\documentclass[11pt, onecolumn]{article}
\usepackage[twoside, left=1in, right=1in, top=1.0in, bottom=1.0in]{geometry}

\usepackage[ntheorem]{empheq}
\usepackage[amsmath,framed,thmmarks,amsthm]{ntheorem}
\usepackage{algorithmicx}
\usepackage{algpseudocode} 
\usepackage{algorithm}

\usepackage{amssymb} % AMS packages
\usepackage[subnum]{cases} % number the formula in cases
\usepackage{graphicx}
\usepackage[svgnames]{xcolor}
\usepackage{enumerate}
\usepackage{caption}
\usepackage{subcaption}
\usepackage{cite}
\usepackage{hyperref}

\usepackage{aliascnt}
\usepackage[capitalize]{cleveref}% must be the last usepackage

\usepackage{nicefrac}

\usepackage{xargs}
\newcommandx{\addpgfplot}[3][1=\textwidth,2=tpdfs]{%
\includegraphics[width=#1]{{#2/#3}.pdf}
}
\newcommand{\CondProb}[2]{\mathbb{P}\left[{#1}\,\middle\vert\,{#2}\right]}
\newcommand{\defeq}{\triangleq}

\newcommand{\eg}{e.g.}

\newcommand{\ie}{i.e.}
\newcommand{\Prob}[1]{\mathbb{P}\left[{#1}\right]}

\newcommand{\nn}{\nonumber}

\newcommand{\tT}{\tilde{T}}
\newcommand{\tC}{\tilde{C}}

\newcommand{\hF}{\hat{F}}
\newcommand{\hx}{\hat{x}}
\newcommand{\hy}{\hat{y}}
\newcommand{\bx}{\mathbf{x}}
\newcommand{\bhx}{\mathbf{\hat{x}}}
\newcommand{\bhy}{\mathbf{\hat{y}}}

\newcommand{\convergeinprob}{\stackrel{P}{\rightarrow}}

\newcommand{\BigO}[1]{O\left({#1}\right)}
\newcommand{\SmallO}[1]{o\left({#1}\right)}

\newcommand{\BigTheta}[1]{\Theta\left({#1}\right)}

\newcommand{\Frechet}{Fr\'{e}chet}

\newcommand{\fccdf}[2][X]{\brF_{{#1}}\left({#2}\right)}
\newcommand{\finvccdf}[2][X]{\brF^{-1}_{#1}\left({#2}\right)}

\newcommand{\ntoinf}[1][n]{{#1}\rightarrow\infty}
\newcommandx{\OSn}[3][1=X,3=n,usedefault]{{#1}_{{#2}:{#3}}}

\newcommand{\E}[1]{\mathbb{E}\left[{#1}\right]}

\newcommand{\iid}{{i.i.d.}}

\newcommand{\Set}[1]{\left\{{#1}\right\}}

\newcommand{\posfunc}[1]{\left({#1}\right)^+}
\newcommand{\beiid}{\stackrel{~\iid}{\sim}}
\newcommand{\SetDef}[2]{\left\{#1 \setst #2 \right\}}

\newcommand{\setst}{:}

\newcommand{\brF}{\bar{F}}
\newcommand{\pdfunc}{{p.d.f.}}
\newcommand{\cdfunc}{{c.d.f.}}

\newcommand{\CDUpper}[1]{\omega\left({#1}\right)}

\newcommand{\reals}{\mathbb{R}}

\newcommand{\EMConstantSymbol}{\gamma_\mathrm{EM}}

\newcommand{\GumbelDist}{\Lambda}
\newcommand{\FrechetDist}{\Phi_\xi}
\newcommand{\RWeibullDist}{\Psi_\xi}

\newcommand{\SingleFork}[1]{\pi\left(#1\right)}
\newcommand{\SingleForkKill}[1]{\pi_{\mathrm{kill}}\left(#1\right)}
\newcommand{\SingleForkKeep}[1]{\pi_{\mathrm{keep}}\left(#1\right)}
\newcommand{\SingleForkKeepText}{\pi_{\mathrm{keep}}}
\newcommand{\SingleForkKillText}{\pi_{\mathrm{kill}}}
\newcommand{\Ykill}{Y_{\mathrm{kill}}}
\newcommand{\Ykeep}{Y_{\mathrm{keep}}}

\newcommand{\fDA}[1]{\mathrm{DA}\left(#1\right)}
\newcommand{\DAG}{\fDA{\Lambda}}
\newcommand{\DAF}[1][\xi]{\fDA{\Phi_{#1}}}
\newcommand{\DAW}[1][\xi]{\fDA{\Psi_{#1}}}

\newcommand{\Tstage}[1]{T^{(#1)}}
\newcommand{\Cstage}[1]{C^{(#1)}}

\newcommand{\Ta}{\Tstage{1}}
\newcommand{\Ca}{\Cstage{1}}
\newcommand{\Tb}{\Tstage{2}}
\newcommand{\Cb}{\Cstage{2}}

\newcommand{\Cost}{C}

\newcommand{\tla}{\tilde{a}}
\newcommand{\tlb}{\tilde{b}}

\newcommand{\EulerConstant}{Euler-Mascheroni constant}
\newcommand{\fGamma}[1]{\Gamma\left({#1}\right)}

\newcommand{\PDPareto}{\textsf{Pareto}}

\newcommand{\PDGaussian}{\textsf{N}}

\newcommand{\PDExp}{\textsf{Exp}}
\newcommand{\PDSExp}{\textsf{ShiftedExp}} % shifted exponential

\newcommand{\PGaussian}[2]{\PDGaussian\left({#1, #2}\right)}

\newcommand{\PPareto}[2]{\PDPareto\left({#1},{#2}\right)}
\newcommand{\PSExp}[2]{\PDSExp\left({#1},{#2}\right)}

\newcommand{\PExp}[1]{\PDExp\left({#1}\right)}

\newtheorem{definition}{Definition}

\newaliascnt{axiom}{theorem}

\aliascntresetthe{axiom}

\newaliascnt{lemma}{theorem}

\aliascntresetthe{lemma}

\newaliascnt{prop}{theorem}

\aliascntresetthe{prop}

\newaliascnt{corollary}{theorem}

\aliascntresetthe{corollary}

\newaliascnt{conjecture}{theorem}

\aliascntresetthe{conjecture}

\newaliascnt{observation}{theorem}

\aliascntresetthe{observation}

\newaliascnt{algo}{theorem}

\aliascntresetthe{algo}

\newaliascnt{notation}{definition}

\aliascntresetthe{notation}

\crefname{equation}{}{}
\Crefname{equation}{}{}
\crefname{thm}{theorem}{theorems}
\Crefname{thm}{Theorem}{Theorems}
\crefname{app}{appendix}{appendices}
\Crefname{app}{Appendix}{Appendices}
\crefname{prop}{proposition}{propositions}
\Crefname{prop}{Proposition}{Propositions}
\crefname{figure}{fig.}{figures}
\Crefname{figure}{Fig.}{Figures}
\crefname{defn}{definition}{definitions}
\Crefname{defn}{Definition}{Definitions}
\crefname{fact}{fact}{facts}
\Crefname{fact}{Fact}{Facts}
\crefname{appendix}{appendix}{appendices}
\Crefname{appendix}{Appendix}{Appendices}
\crefname{algo}{algorithm}{algorithms}
\Crefname{algo}{Algorithm}{Algorithms}
\crefname{algorithm}{algorithm}{algorithms}
\Crefname{algorithm}{Algorithm}{Algorithms}
\crefname{conjecture}{conjecture}{conjectures}
\Crefname{conjecture}{Conjecture}{Conjectures}
\crefname{obs}{observation}{observations}
\Crefname{obs}{Observation}{Observations}

\crefformat{equation}{(#2#1#3)}

\newtheorem{lem}{Lemma}
\newtheorem{thm}{Theorem}
\newtheorem{defn}{Definition}
\newtheorem{coro}{Corollary}

\newtheorem{rem}{Remark}

\crefname{equation}{}{}
\Crefname{equation}{}{}
\crefname{thm}{theorem}{theorems}
\Crefname{thm}{Theorem}{Theorems}
\crefname{clm}{claim}{claims}
\Crefname{clm}{Claim}{Claims}
\Crefname{coro}{Corollary}{Corollaries}
\Crefname{lem}{Lemma}{Lemmas}
\Crefname{sec}{Section}{Sections}
\crefname{app}{appendix}{appendices}
\Crefname{app}{Appendix}{Appendices}
\Crefname{part}{Part}{Parts}
\crefname{prop}{proposition}{propositions}
\Crefname{prop}{Proposition}{Propositions}
\Crefname{propty}{Property}{Properties}
\crefname{figure}{fig.}{figures}
\Crefname{figure}{Fig.}{Figures}
\crefname{defn}{definition}{definitions}
\Crefname{defn}{Definition}{Definitions}
\crefname{fact}{fact}{facts}
\Crefname{fact}{Fact}{Facts}
\crefname{appendix}{appendix}{appendices}
\Crefname{appendix}{Appendix}{Appendices}
\crefname{algo}{algorithm}{algorithms}
\Crefname{algo}{Algorithm}{Algorithms}
\crefname{algorithm}{algorithm}{algorithms}
\Crefname{algorithm}{Algorithm}{Algorithms}
\crefname{conj}{conjecture}{conjectures}
\Crefname{conj}{Conjecture}{Conjectures}
\crefname{obs}{observation}{observations}
\Crefname{obs}{Observation}{Observations}
\crefname{rem}{remark}{remarks}
\Crefname{rem}{Remark}{Remarks}

\def\mytitle{Efficient Straggler Replication in Large-scale Parallel Computing}

\DeclareGraphicsExtensions{.pdf,.png,.jpg}

\begin{document}
\title{\mytitle
\thanks{
This work was supported, in part, 
by NSF under Grant No.~CCF-1319828, 
AFOSR under Grant No.~FA9550-11-1-0183,
Wellington and Irene Loh Fund Fellowship,
Schlumberger Foundation Faculty for the Future Fellowship, and 
Claude E. Shannon Research Assistantship.
We thank Devavrat Shah for helpful discussions.
}
}
\date{}

\author{
Da Wang
\thanks{
Signals, Information and Algorithms Laboratory, Massachusetts Institute of Technology, Cambridge, MA, 02139 
}
\\
\texttt{dawang@alum.mit.edu}
\and
Gauri Joshi
\thanks{Department of Electrical and Computer Engineering, Carnegie Mellon University, Pittsburgh, PA, 15213}
\\
\texttt{gaurij@andrew.cmu.edu}
\and
Gregory Wornell\footnotemark[2]
\\
\texttt{gww@mit.edu}
}

\maketitle

\begin{abstract}
    In a cloud computing job with many parallel tasks, the tasks on the slowest machines (straggling tasks) become the bottleneck in the job completion. Computing frameworks such as MapReduce and Spark tackle this by replicating the straggling tasks and waiting for any one copy to finish. Despite being adopted in practice, there is little analysis of how replication affects the latency and the cost of additional computing resources.  In this paper we provide a framework to analyze this latency-cost trade-off and find the best replication strategy by answering design questions such as: 1) when to replicate straggling tasks, 2) how many replicas to launch, and 3) whether to kill the original copy or not. Our analysis reveals that for certain execution time distributions, a small amount of task replication can drastically reduce both latency as well as the cost of computing resources. We also propose an algorithm to estimate the latency and cost based on the empirical distribution of task execution time. Evaluations using samples in the Google Cluster Trace suggest further latency and cost reduction compared to the existing replication strategy used in MapReduce.

\end{abstract}

\section{Introduction}
%Recently cloud computing is being offered as a service by Amazon EC2, Microsoft Azure etc.\ , where users can rent machines by the hour to run computing jobs. 
In cloud computing, large-scale sharing of computing resources provides users with great flexiblity and scalability. Computing frameworks such as MapReduce~\cite{map_reduce} and Apache Spark~\cite{zaharia_spark_2010} are developed to harness these benefits. These
frameworks employ massive parallelization by dividing a large job into many tasks that can be executed parallely on different machines. These frameworks can be used to run optimization and machine learning algorithms that can be easily divided into independent parallel tasks, for example alternating direction method of multipliers (ADMM)~\cite{boyd_distributed_2011} and Markov Chain Monte-Carlo (MCMC)~\cite{neiswanger_asymptotically_2013}.

The execution time of a task on a machine is subject to stochastic variations due to co-hosting, virtualization and other hardware and network variations~\cite{dean_tail_2013}. Thus, a key challenge in executing a job that
consists of a large number of parallel tasks is the latency in waiting for the slowest tasks, or the ``stragglers'' to finish. As pointed out in \cite[Table 1]{dean_tail_2013}, the latency of executing many parallel tasks could be significantly larger ($140$ ms) than the median latency of a single task ($1$ ms).

In this work we provide a mathematical framework to analyze how replication of straggling tasks affects the latency and the cost of computing resources, and propose better scheduling policy designs.

\subsection{Related prior work}
\label{subsec:prior_work}
The idea of replicating tasks in parallel computing has been recognized by system
designers~\cite{ghare_improving_2005}, and first adopted at a large scale via the ``backup tasks'' in MapReduce~\cite{map_reduce}. 
A line of systems work \cite{zaharia_rdd_2012,zaharia_sparrow_2013,ananthanarayanan_effective_2013} and references therein further developed this idea.  For example, Apache Spark implements ``speculative execution'' to allow relaunching slow running tasks. 

While task replication has been studied in systems literature and also adopted in practice, there is not much work on mathematical analysis of replication strategies. Replication strategies are analyzed in \cite{wang_efficient_2014}, mainly for the single task case. In this paper we consider task replication for a
job consisting of a large number of tasks, which corresponds more closely to today's large-scale cloud computing
frameworks.

The use of redundancy to reduce latency has also attracted attention in other contexts such as cloud storage and
networking~\cite{gauri_yanpei_emina_jsac,vulimiri_low_2013,gauri_allerton_2015,sun_shroff_2015,mds_queue,gardner_sigmetrics_2015}.
Most of these works that consider queueing focus on the case of one task. Waiting for many tasks is harder to analyze as indicated by fork-join queue analysis.

\subsection{Our contributions}
In this work we propose a framework to analyze strategies for replicating straggling tasks of a large computing job. In particular we consider three parameters of a straggler replication strategy: 1) the fraction of tasks declared as stragglers, 2) number of replicas for each straggling tasks, and 3) whether the original copy should be killed or kept running. We characterize how these parameters impact the trade-off between latency and computing cost. Our characterizations allow us to identify regimes with the surprising property that
%Surprisingly, for heavy-tail distributions we can identify regimes where
replicating a small fraction of tasks drastically reduces latency while saving computing cost. 
%We present both analytical solutions for common distributions such as Pareto and Exponential, and algorithms for arbitrary execution time distributions. 
These insights allow one to apply optimization to search for scheduling policies
based on one's sensitivity to computing latency and computing cost.

%We show that the tail distribution of the task service time is the key factor in characterizing the scaling of latency with respect to the number of tasks. 

%The choice of a straggler replication strategy involves optimizing the following:
%\begin{itemize}
%\item When to replicate: the fraction of tasks to declare as stragglers and replicate
%\item How many replicas to launch
%\item Whether to kill original straggling copies or not
%\end{itemize}
%
%In particular for heavy tail distributions such as the Pareto distribution, we identify scenarios where the
%latency and computing cost can be reduced simultaneously. 
%For new-shorter-than-used distributions, we show that it is better to kill original copies of tasks.  
%%We also propose a resampling good task replication policy when it is difficult to use
%%the proposed analysis techniques for the empirical distribution of task execution time. 
%In addition, we propose computational methods for deriving the tradeoff between latency and computing resources
%based on the empirical distribution of task execution time. 

%\TODO{make the above summary better}

%\subsection{Organization}
The rest of the paper is organized as follows. 
In \Cref{sec:scheduling_formulation} we introduce notation, formulate the problem, and define performance
metrics used in the paper.  
In \Cref{sec:single_fork} we provide an analysis of single-fork task replication
policies and defer all proofs to \Cref{sec:appendix}.
Then in \Cref{sec:heuristic} we describe an algorithm that finds a good scheduling policy for execution time distributions that are not analytically tractable (\eg,
empirical distributions from real-world traces). In \Cref{sec:conc_remarks} we conclude with a discussion of the
implications and future perspectives.

\section{Problem Formulation}
\label{sec:scheduling_formulation}
%We now describe the system model and the performance metrics used to evaluate replication
%strategies. 

\subsection{Notation}
\label{sec:scheduling_notation}
Lower-case letters (\eg, $x$) denote a particular value of the corresponding random variable, which is denoted in upper-case letters (\eg, $X$). We denote the cumulative
distribution function (\cdfunc) of $X$ by $F_X(x)$. Its complement, the tail distribution is denoted by $\brF_X(x) \defeq 1 - F_X(x)$. We denote the upper end point of $F_X$ by
\begin{equation}
    \CDUpper{F_X} \defeq \sup \SetDef{x}{F_X(x) < 1} .
    \label{eq:def_support}
\end{equation}

For \iid\ random variables $X_1, X_2, \cdots, X_n$, we define $X_{j:n}$ as the $j$-th order statistic, \ie, the $j$-th smallest of the $n$ random variables.

\subsection{System Model}
\label{sec:sys_model}
We consider a job consisting of $n$ \emph{parallel tasks}, where $n$ is large\footnote{Analysis of real-world trace data shows that it is common for a job to contain hundreds or even thousands of tasks~\cite{reiss_towards_2012}.} and each task is assigned to a different machine.
We use the probability distribution $F_X$ to model the random variation in machine response time due to factors such as
congestion, queueing, virtualization, and competing jobs being run on the same machines, and assume this
execution time distribution is independent and identically distributed (\iid) across machines.
The \emph{identical} assumption of $F_X$ implies that
tasks in this job are assigned to machines with processing power proportional to task size, with the simplest case being
a group of homogeneous tasks are assigned to a group of homogeneous machines.
The \emph{independent}
assumption of $F_X$ could be satisfied when machine response times fluctuate independently over time, or when
each new task (or new replica) is assigned to a new machine that is not previously used to run tasks of the
current job. Note that we treat the variability that $F_X$ captures as an exogenous factor from a user's perspective---in general a user renting machines from a cloud computing service has little or no control
over other jobs that share the resources.%
\footnote{%
A system designer may be able to influence this variability by adjusting the resource sharing among different
jobs, another interesting direction that is beyond the scope of this work.}

\subsection{Scheduling Policy}
A \emph{scheduling policy} or \emph{scheduler} assigns one or more replicas of each task to different machines,
possibly at different time instants. In this work, we assume the scheduler receives instantaneous feedback notifying it
when a machine finishes its assigned task, and there is \emph{no intermediate feedback} indicating the status of
processing of a task. 
We focus our attention on a set of policies called \emph{single-fork policies}, defined as follows.

\begin{figure}
\centering
\begin{subfigure}[t]{0.42\textwidth}
    \centering
    \includegraphics[width=0.92\linewidth ]{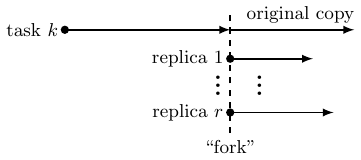}
	\caption{Keep the original copy ($\SingleForkKeepText$)}
\end{subfigure}
~
\begin{subfigure}[t]{0.42\textwidth}
    \centering
    \includegraphics[width=0.92 \linewidth]{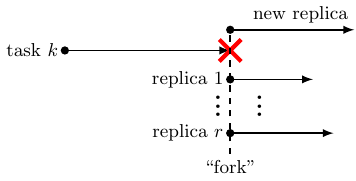}
\caption{Kill the original copy ($\SingleForkKillText$)}
\end{subfigure}
\caption{Single-fork policy illustration}
\label{fig:single_fork_relaunching}
\end{figure}

\begin{definition}[Single-fork scheduling policy]
\label{defn:single_fork}
A {single-fork scheduling policy} $\SingleFork{p,r}$ launches all $n$ tasks at time $0$. It waits until
$(1-p)n$ tasks finish. For each of the remaining $pn$ straggling tasks, it chooses one of the following two actions: 
\begin{itemize}
    \item \textbf{replicate and keep the original copy ($\SingleForkKeep{p,r}$):} launch $r$ new replicas;
    \item \textbf{replicate and kill the original copy ($\SingleForkKill{p,r}$):} kill the original copy and launch $r+1$ new replicas.
\end{itemize}
When the earliest replica of a task finishes, all the other remaining replicas of the same task are terminated.
\end{definition}
\medskip
%We use $\SingleForkKeep{p,r}$ and $\SingleForkKill{p,r}$ to distinguish the cases that we keep or kill the original copy. 
%We use $l$ to denote the number of original replicas of each task remaining after the forking point. Hence $l=0$
%when the original replica is killed and restarted, and $l=1$ otherwise. 
Note that in both scenarios there are a total of $r+1$ replicas running after the forking point. \Cref{fig:single_fork_relaunching} illustrates these two cases of keeping or killing the original copy of a task. For simplicity of notation we assume that $p$ is such that $pn$ is an integer. We note that $p=0$ corresponds to
running $n$ tasks in parallel and waiting for all to finish, which is the baseline case without any replication or killing any original tasks.

\begin{rem}[Backup tasks in MapReduce and Spark] 
\label{rem:single_fork_mapreduce}
The idea of ``backup tasks'' in Google's MapReduce~\cite{map_reduce}, and ``speculative execution'' in Apache Spark~\cite{zaharia_spark_2010}  corresponds to a single-fork policy with $r=1$ and $\SingleForkKeepText$. The value of $p$ is tuned dynamically and hence not specified in~\cite{map_reduce}. The \texttt{spark.speculation.quantile} configuration in Apache Spark corresponds to $p$ in the single-fork policy.
\end{rem}

Although we focus on single-fork policies in this paper, the analysis can be generalized to multi-fork policies,
where new replicas of straggling tasks are launched at multiple times during the execution of the
job~\cite[Section 6.4]{wang_thesis_2014}. Forking multiple times can achieve a better latency-cost trade-off, but could be
undesirable in practice due to additional delay and complexity in obtaining new and killing existing replicas. 

\subsection{Performance Metrics}
\label{sec:scheduling_metrics}
We now define the latency and cost metrics used to compare straggler replication policies and understand when and how replication is useful.
\begin{defn}[Expected latency]
\label{defn:latency}
Given a scheduling policy, the expected latency $\E{T}$ is the expected value of $T$, the time taken for at least one replica of each of the $n$ tasks to finish. It can be expressed as
\begin{align}
\E{T} &= \mathbb{E} \left[ \max_{i \in \{1, 2, \dots, n\} } T_i \right], \label{eq:latency}
\end{align}
where $T_i$ is the time when at least one replica of task $i$ finishes. 
More specifically, suppose the scheduler launches $r$ replicas of each of the $n$ tasks at times $t_{i,j}$ for $j = 0, 1, 2, \dots r$, then
\begin{align}
  T_i &= \min_{0 \leq j \leq r} (t_{i,j} + X_{i,j}), 
  \label{eq:task_latency}
\end{align}
where $X_{i,j}$ are i.i.d., drawn from the execution time distribution $F_X$.  
\end{defn}
\begin{defn}[Expected cost]
\label{defn:cost}
The expected computing cost $\E{\Cost}$ is the sum of the running times of all machines, normalized by $n$, the
number of tasks in the job. The running time is the time from when the task is launched on a machine, until it
finishes, or is killed by the scheduler. 
More specifically, suppose the scheduler launches $r$ replicas of each of the $n$ tasks at times $t_{i,j}$ for $j = 0, 1, 2, \dots r$, then
\begin{align}
    \Cost &\defeq \frac{1}{n} \sum_{i=1}^n \sum_{j=0}^{r} \posfunc{ T_i - t_{i,j} },
    \label{eq:cost_cloud}
\end{align}
where 
$T_i$ is given in \Cref{eq:task_latency} and $\posfunc{x} = \max(0,x)$.
\end{defn}

%\begin{rem}
%\label{rem:cost}
Infrastructure as a Service (IaaS) providers such as Amazon Web Services (AWS), Microsoft Azure, and Google Cloud Platform charge users by the time and the number of machines used. Then the money spent by a user to rent the machines is
proportional to our cost metric $\E{\Cost}$.
%\end{rem}

\begin{figure}[t]
    \begin{center}
        \includegraphics[width=0.5\linewidth]{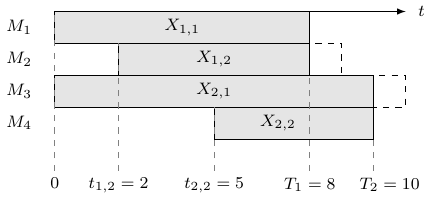}
    \end{center}
    \caption{Illustration of $T$ and $C$ for a job with two tasks, and two replicas of each task. The latency $T = \max(8,10) = 10$, and the computing cost is $C = (8+6+10+5)/2 = 14.5$.}
    \label{fig:multi_task_scheduling}
\end{figure}

\Cref{fig:multi_task_scheduling} illustrates the execution of a job with two tasks, and evaluation of the corresponding latency $T$ and cost $C$. Given two tasks, we launch two replicas of task 1 $t_{1,1} = 0$ and $t_{1,2} = 2$, and two replicas of task 2 at $t_{2,1} = 0$ and $t_{2,2} = 5$. The task execution times are
$X_{1,1} = 8$, $X_{1,2} = 7$, $X_{2,1} = 11$, and $X_{2,2} = 5$. Machine $M_1$ finishes the task first at time
$t=8$, $T_1 = 8$ and the second replica running on $M_2$ is terminated before it finishes executing. Similarly,
machine $M_4$ finishes task $2$ at time $T_2 = 10$, and the replica running on $M_3$ is terminated. Thus the
latency of the job is $T = \max\Set{T_1, T_2} = 10$. The cost is the sum of all running times normalized by $n$,
\ie, $\Cost = (8 + 6 +  10 + 5)/2 = 14.5$.
%

%\begin{definition}[Sub-optimal single-fork policies]
%We say a policy $\SingleFork{p,r,l}$ is \emph{sub-optimal} if there exists another single-fork policy
%$\SingleFork{p',r,l}$ which achieves lower $\E{T}$ and $\E{C}$ than $\SingleFork{p,r,l}$.
%\end{definition}

\section{Single-fork policy analysis}
\label{sec:single_fork}
In this section we analyze the trade-off between the performance metrics $\E{T}$ and $\E{C}$ for the single-fork policy defined in \Cref{defn:single_fork}. The choice of the best single fork policy depends on the tail of $F_X$, as we demonstrate for the Shifted exponential and Pareto distributions. All proofs are deferred to \Cref{sec:appendix}.

\subsection{Performance characterization}
\label{sec:single_fork_calcs}

\begin{thm}[Single-Fork Latency and Cost]
\label{thm:single_fork_gen}
For a computing job with $n$ tasks, and task execution time distribution $F_X$, the latency and cost metrics as $n \rightarrow \infty$ are
\begin{align}
\E{T} &= F_X^{-1}(1-p) + \E{Y_{pn:pn}} \label{eqn:latency_gen},\\
\E{C} &= \int_{0}^{1-p} F_X^{-1}(h) dh + p F_X^{-1}(1-p)  + (r+1) p \cdot \E{Y}, \label{eqn:cost_gen}
\end{align}
where $Y$ is the residual execution time of a straggling tasks after launching replicas. Its 
tail distribution $\bar{F}_Y$ is given by
\begin{align}
\fccdf[Y]{y} = 
\begin{cases}
\fccdf[X]{y}^{r+1} & \text{for } \SingleForkKill{p,r} ,\\
\frac{1}{p} \fccdf[X]{y}^r \fccdf[X]{y+F_X^{-1}(1-p)} & \text{for } \SingleForkKeep{p,r}.
\end{cases} 
\label{eqn:Y_defn}
\end{align}
The second term $\E{Y_{pn:pn}}$ in \eqref{eqn:latency_gen} is the expected maximum of $pn$ i.i.d.\ random variables drawn from $F_Y$. Its behavior as $n \rightarrow \infty$ is given by the Extreme Value Theorem (\Cref{thm:ev_thm}).
\end{thm}

The proof of \Cref{thm:single_fork_gen} can be found in \Cref{sec:appendix}. A key observation from \Cref{thm:single_fork_gen} is that the execution time before forking, $F_X^{-1}(1-p)$, is a quantity \emph{independent with respect to $n$} and monotonically non-increasing with $p$, while the execution time after forking, $\E{Y_{pn:pn}}$, is monotonically non-decreasing with $pn$. In certain regimes, increasing $p$ (and with proper choice of $r$), the time reduction in first stage outweighs the time increase in the second stage, reducing the overall execution latency. %For example, if we kill the original copy and choose $r=2$, the tail of distribution $\bar{F}_Y = \bar{F}_X^2$, because two identical replicas with distribution $F_X$ are launched at the forking point. 

Using \Cref{thm:single_fork_gen} we can determine the single-fork policy parameters $p$ and $r$ that give the best latency-cost trade-off for a given service time distribution $F_X$. To decide whether to kill or to keep the original copy of the straggling task, we are essentially comparing the additional time needed for the original time to finish and the completion time for a new copy. In \Cref{lem:relaunch_vs_not} we identify when killing the original task is better than keeping the original task and vice versa.

\begin{lem}[Kill or keep original task]
\label{lem:relaunch_vs_not}
  For a given $0 < p \leq 1$, killing the original task gives lower latency and cost than keeping it running if
    \begin{align}
    \frac{1}{p}\Pr(X>x+F_X^{-1}(1-p)) \geq \Pr(X>x) \quad \text{for all } x \geq 0.
    \label{eqn:kill_or_keep}
    \end{align}
    Conversely, if the inequality in \eqref{eqn:kill_or_keep} is reversed for all $x \geq 0$, then keeping the original task is better.
\end{lem}

The proof is given in \Cref{sec:appendix}. For a class of distributions called `new-longer-than-used' distributions \cite{kochar_partial_1987}, \eqref{eqn:kill_or_keep} is true for any $0 < p \leq 1$. An example of such distributions is the shifted-exponential distribution for which we analyze the latency-cost trade-off in \Cref{sec:single_fork_egs} below.

\subsection{Single-fork scheduling with analytical execution time distributions}
\label{sec:single_fork_egs}
%By \Cref{thm:single_fork_gen} the scaling of $\E{Y_{pn:pn}}$ with $n$ depends on whether the task service time $X$ is heavy, light or exponential tailed. And by \Cref{lem:relaunch_vs_not}, we know that the choice between $\SingleForkKillText$ or $\SingleForkKeepText$ is governed by whether $X$ is new-longer-than-used or new-shorter-than used. 

In this section we evaluate the latency-cost trade-off in \Cref{thm:single_fork_gen} for two execution time
distributions: Shifted exponential and Pareto. The shifted exponential distribution has an exponential tail,
while Pareto distribution has a heavy tail. 

%Light-tailed distributions generally do not benefit from task replication and thus we do not analyze a
%light-tailed example here. %The shifted exponential distribution is new-longer-than-used. The tail $\Pr(X>x)$
%of the Pareto distribution is new-shorter-than-used for $x \geq x_m$, but not otherwise. %Thus, Pareto is
%neither new-longer-than-used nor new-shorter-than-used. For these two distributions we demonstrate how the
%tail, and the residual life of the service time distribution affect the choice of the best single-fork policy. 

\subsubsection{Shifted exponential execution time}
\label{sec:single_fork_exp}
Consider that the task execution time distribution $F_X$ is a \emph{shifted exponential distribution} $\PSExp{\Delta}{\mu}$. Its tail distribution function is given by
\begin{align}
    \label{eqn:shifted_exp_dist}
    \Pr(X>x) = \begin{cases}
     e^{-\mu (x-\Delta )}& \text{for } x \geq \Delta , \\
     1 & \text{otherwise.}
     \end{cases}
\end{align}
The shifted exponential distribution has an exponentially decaying tail. 
It is lower bounded by a constant $\Delta$, aiming to capture the delay due to machine start-up or task initialization. Due to this constant $\Delta$, the shifted exponential distribution satisfies \eqref{eqn:kill_or_keep} for any $0 < p \leq 1$. Thus, it is always better to keep the original straggling task, and launch additional replicas if necessary. %new-longer-than-used. The special case $\Delta = 0$ corresponds to the pure exponential distribution, which is both new-longer-than-used and new-shorter-than-used, which implies that it is memoryless.
%But it may not be suitable for modeling the  execution time of a task, as a task seldom finishes instantaneously, and usually 
%Hence we add the constant delay $\Delta$ and model the execution time using the $\PSExp{\Delta}{\mu}$. 
%
\begin{thm}
\label{thm:single_fork_sexp}
For a computing job with $n$ tasks, if the execution time distribution of tasks are \iid\ 
$\PSExp{\Delta}{\mu}$, then as $n \rightarrow \infty$, the latency and cost metrics are
\begin{align}
\E{T} &= 
    \begin{cases}
        \frac{2r +1}{r+1}\Delta + \frac{1}{(r+1)\mu} \left( \ln n - r \ln p + \EMConstantSymbol \right)
        &
        \text{for } \SingleForkKeep{p,r}
        \\
        2\Delta + \frac{1}{(r+1)\mu} \left( \ln n - r \ln p + \EMConstantSymbol \right)
        &
        \text{for } \SingleForkKill{p,r}
    \end{cases}
    ,
\\
\E{\Cost}
    &=
    \begin{cases}
    \Delta + \frac{1}{\mu} + p \left[ 
        \Delta + r  \frac{\left(1 - e^{-\mu \Delta}\right)}{\mu}
    \right]
    &
    \text{for } \SingleForkKeep{p,r}
    \\
    \Delta+ \frac{1}{\mu} + p (r+2) \Delta
    &
    \text{for } \SingleForkKill{p,r}
    \end{cases}
    ,
\end{align}
where $\EMConstantSymbol$ is the Euler-Mascheroni constant,
\begin{align}
\gamma &\defeq \int_1^\infty\left(\frac1{\lfloor x\rfloor}-\frac1{x}\right)\,dx \approx 0.577, \label{eqn:EM_constant_def}
\end{align}
\end{thm}

\begin{figure}[t]
    \centering
    \hspace{-0.3em}
    \addpgfplot[0.45\linewidth]{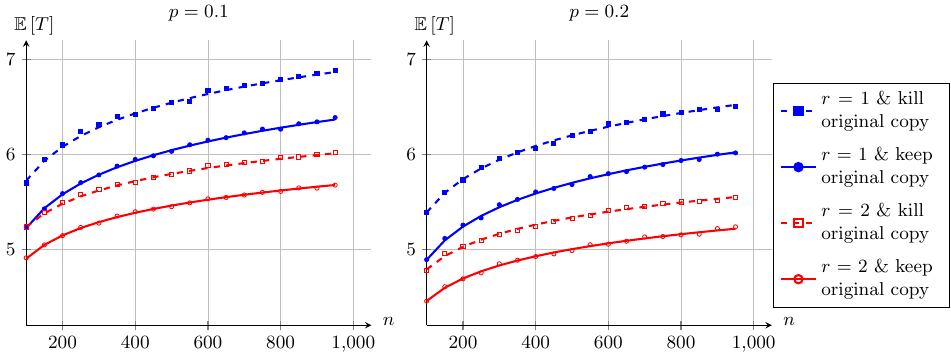}
    \caption{Comparison of the expected latency $\E{T}$ obtained from simulation (points) and analytical
    calculations (lines) for the shifted exponential distribution $\PSExp{1}{1}$.}
    \label{fig:sexp_sim_calc_cmp}
\end{figure}

\begin{figure}
    \begin{subfigure}[b]{\textwidth}
        \centering
        \addpgfplot[\linewidth]{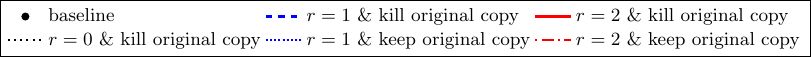}
    \end{subfigure}
    \begin{subfigure}[b]{0.33\textwidth}
        \centering
        \addpgfplot[\linewidth]{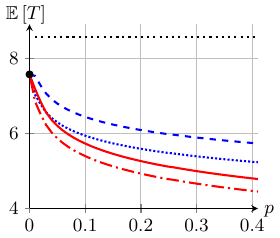}
        \captionsetup{format=hang}
        \caption{Expected latency $\E{T}$}
        \label{fig:exp_ET_vs_p}
    \end{subfigure}
    \begin{subfigure}[b]{0.33\textwidth}
        \centering
        \addpgfplot[\linewidth]{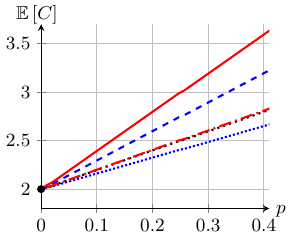}
        \captionsetup{format=hang}
        \caption{Expected cost $\E{\Cost}$}
        \label{fig:exp_EC_vs_p}
    \end{subfigure}%
    \begin{subfigure}[b]{0.33\textwidth}
        \centering
        \addpgfplot[\linewidth]{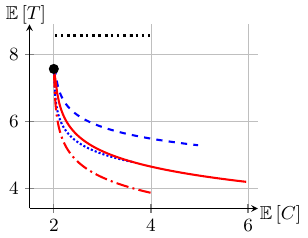}
        \captionsetup{format=hang}
        \caption{Trade-off between $\E{T}$ and $\E{\Cost}$}
        \label{fig:exp_ET_and_ECloud}
    \end{subfigure}
    \caption{Characterization for $\PSExp{1}{1}$ and $n=400$, by varying $p$ in the range of $[0.05, 0.95]$.}
    \label{fig:exp_combined}
\end{figure}

The proof is given in \Cref{sec:appendix}. \Cref{fig:sexp_sim_calc_cmp} compares the latency obtained from Monte-Carlo simulation and analytical calculations for the shifted exponential distribution, indicating that the latency obtained from analytical calculation is very close to the simulated performance for $n\geq 100$, especially for the case with killing the original task.
From \Cref{thm:single_fork_sexp} we observe that given $r$ and whether we kill or keep the original task, replicating earlier (larger $p$) gives an $\Theta(\ln p)$ decrease in latency, and a linear increase the cost. This is also illustrated in \Cref{fig:exp_ET_vs_p,fig:exp_EC_vs_p} for execution time distribution $\PSExp{1}{1}$ and $n=400$. \Cref{fig:exp_ET_and_ECloud} illustrates the latency-cost trade-off. For the special case of $\Delta = 0$ by \Cref{thm:single_fork_sexp}, the cost $\E{C} = 1/\mu$, which is independent of $p$ and $r$. But latency always reduces with $r$ and $p$. This suggests that we can achieve arbitrarily low latency without any increase in cost. However, in practice the minimum time to complete a task is strictly positive, that is $\Delta > 0$. %Thus, pure exponential task service time is not a useful model for the purpose of analyzing task replication.

\subsubsection{Pareto execution time}
\label{sec:single_fork_pareto}
The tail distribution function of the Pareto distribution $\PPareto{\alpha}{x_m}$ is
\begin{equation}
    \label{eqn:pareto_dist}
 \Pr(X> x)
    \defeq
    \begin{cases}
        \left( \frac{x_m}{x} \right)^\alpha & x \geq x_m,
        \\
        1 & \text{otherwise}
    \end{cases}
\end{equation}
The Pareto distribution has a heavy-tail that decays polynomially. 
It has been observed to fit task execution time distributions in data centers~\cite{reiss_towards_2012,dean_tail_2013}. 

\begin{figure}
    \centering
    \hspace{-0.3em}
    \addpgfplot[0.45\linewidth]{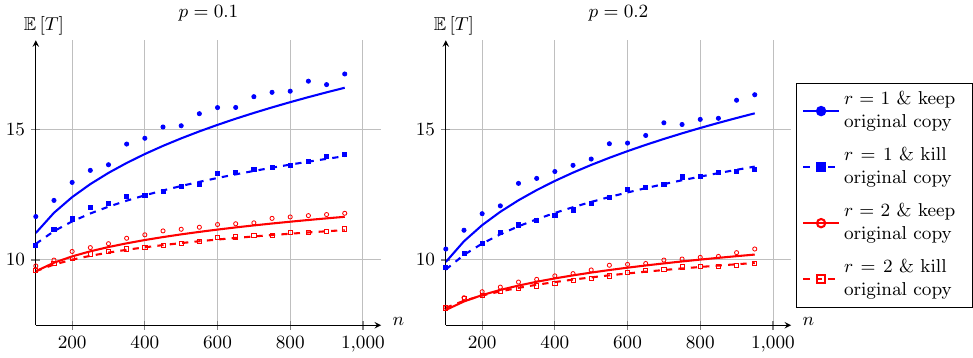}
    \caption{Comparison of the expected latency $\E{T}$ obtained from simulation (points) and analytical
    calculations (lines) for the Pareto distribution $\PPareto{2}{2}$.}
    \label{fig:pareto_sim_calc_cmp}
\end{figure}

\begin{thm}
\label{thm:single_fork_pareto}
For a computing job with $n$ tasks, if the execution time distribution of tasks are \iid\ $\PPareto{\alpha}{x_m}$, then as $n \rightarrow \infty$, 
the latency and cost metrics are
\begin{align}
    \label{eqn:pareto_ET}
    \E{T} &= x_m p^{-1/\alpha} + \fGamma{1 - \frac{1}{(r+1)\alpha}} \tla_{pn},\\
\E{C} &= x_m \frac{\alpha}{\alpha-1} - x_m \frac{p^{1- 1/\alpha}}{\alpha-1} + (r+1) p \cdot \E{Y}.
\end{align}
The values of $\tla_{pn}$ and $\E{Y}$ depend on the whether we choose to keep or kill the original task, and are given as follows.

~\\%
\textbf{Case 1: Killing the original task} 
\begin{align}
    \tla_{pn} &= (pn)^{\frac{1}{(r+1)\alpha}} x_m,\label{eqn:pareto_relaunching_a}\\
    \E{Y} &= \frac{(r+1)\alpha}{(r+1)\alpha-1} x_m .
\end{align}
\textbf{Case 2: Keeping the original task}\\
The tail distribution of $Y$
\begin{align}
\bar{F}_Y(y) &= \frac{1}{p} \left( \frac{x_m}{y} \right)^{\alpha r} \left( \frac{x_m}{y + x_m p^{-1/\alpha}} \right)^{\alpha} .\label{eqn:Y_pareto_no_relaunch_first}
\end{align}
The term $\tilde{a}_{pn} = \bar{F}_Y^{-1}\left(\frac{1}{pn} \right)$, and $\E{Y}$ can be evaluated numerically by integrating \eqref{eqn:Y_pareto_no_relaunch_first} from $y = 0$ to $\infty$.
%
%The term $\tla_{pn}$ is the solution to
%\begin{align}
%    \label{eqn:tla_n_for_pareto_no_relaunching}
%    n^{1/\alpha} x_m^{r+1} &=  x_m p^{-1/\alpha}\tla_{pn}  ^r + \tla_{pn} ^{r+1}.
%\end{align}
%and $\E{Y}$ is evaluated numerically as discussed in the proof.
\end{thm}

The proof is given in \Cref{sec:appendix}. Similar to \Cref{fig:sexp_sim_calc_cmp}, \Cref{fig:pareto_sim_calc_cmp} compares the latency obtained from simulation and analytical calculations for the Pareto distribution, which again demonstrates the effectiveness of the asymptotic theory. Based on \Cref{thm:single_fork_pareto}, we can derive how $\E{T}$ scales with $n$ in the following corollary.
\begin{coro}
\label{coro:scaling}
For a computing job with $n$ tasks, if the execution time distribution of each task is $\PPareto{\alpha}{x_m}$, then the expected latency satisfies 
$$
\E{T} = \BigTheta{n^{1/(\alpha(r+1))}}.
$$
\end{coro}
%The proofs of both \Cref{theorem:single_fork_pareto} and \Cref{coro:scaling} are given in \Cref{sec:proof_single_fork}. 
\Cref{coro:scaling} indicates that the heavier the tail (smaller $\alpha$), the faster $\E{T}$ grows with $n$. We also observe that the latency reduction due to redundancy $r$ diminishes as $r$ increases due to the $1/(r+1)$ factor in the exponent.
%\begin{figure}[bt]
%    \centering
%    \hspace{-0.3em}
%    \addpgfplot[0.43\linewidth]{pareto_ET_and_EC_vs_p_n400.tikz}
%    \caption{Expected latency and cloud user cost for a Pareto execution time
%    distribution $\PPareto{2}{2}$, given $n=400$.}
%    \label{fig:pareto_ET_and_EC_vs_p}
%\end{figure}

\begin{figure}
    \begin{subfigure}[b]{\textwidth}
        \centering
        \addpgfplot[\linewidth]{tradeoff_legend.tikz}
    \end{subfigure}
    \begin{subfigure}[b]{0.33\textwidth}
        \centering
        \addpgfplot[\linewidth]{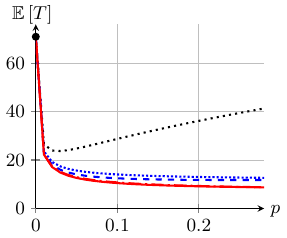}
        \captionsetup{format=hang}
        \caption{Expected latency $\E{T}$}
        \label{fig:pareto_ET_vs_p}
    \end{subfigure}
    \begin{subfigure}[b]{0.33\textwidth}
        \centering
        \addpgfplot[\linewidth]{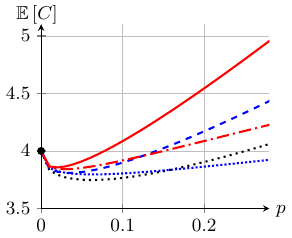}
        \captionsetup{format=hang}
        \caption{Expected cost $\E{\Cost}$}
        \label{fig:pareto_EC_vs_p}
    \end{subfigure}%
    \begin{subfigure}[b]{0.33\textwidth}
        \centering
        \addpgfplot[\linewidth]{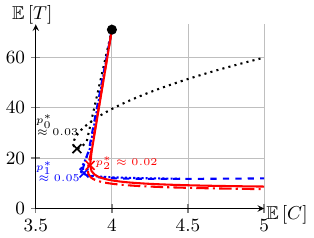}
        \captionsetup{format=hang}
        \caption{Trade-off between $\E{T}$ and $\E{\Cost}$}
        \label{fig:pareto_ET_and_ECloud}
    \end{subfigure}
    \caption{Characterization for $\PPareto{2}{2}$ and $n=400$, by varying $p$ in the range of $[0.05, 0.95]$.}
    \label{fig:pareto_combined}
\end{figure}

In \Cref{fig:pareto_ET_vs_p,fig:pareto_EC_vs_p} we plot the expected latency and cost as $p$ varies, for different values of $r$. The black dot is the baseline case ($p=0$), where no replication is used and we simply wait for the original copies of all $n$ tasks to finish. Note that $r = 0$ and keeping the original copy is also equivalent to the baseline case, and thus not plotted in the figures. The diminishing return of increasing $r$ in terms of latency reduction is clearly demonstrated. In addition, we observe that a small amount of replication (small $p$ and $r$) can reduce latency significantly in comparison with the baseline case. But as $p$ increases further, the latency may increase (as observed for $r=0$) because of the second term in \Cref{eqn:latency_gen}. 

%\begin{figure}[bt]
%    \centering
%    \hspace{-0.3em}
%    \addpgfplot[0.43\linewidth]{pareto_ET_and_EC_vs_p_n400.tikz}
%    \caption{Expected latency and cloud user cost for a Pareto execution time
%    distribution $\PPareto{2}{2}$, given $n=400$.}
%    \label{fig:pareto_ET_and_EC_vs_p}
%\end{figure}

Intuition suggests that replicating earlier (larger $p$) and more (higher $r$) will increase the cost $\E{C}$. But \Cref{fig:pareto_ET_vs_p,fig:pareto_EC_vs_p} show that this is not necessarily true. Since we kill replicas of task when one of its replicas finish, there could in fact be a saving in the computing cost. However
this benefit diminishes as $p$ and $r$ increase above a certain threshold.

\Cref{fig:pareto_ET_and_ECloud} shows the latency versus the computing cost for different values of $r$, with $p$ varying along each curve. Depending upon the latency requirement and limit on the cost, one can choose
an appropriate operating point on this trade-off curve. This plot again demonstrates the non-intuitive phenomenon that it is possible to reduce latency (from $70$ to about $15$ for $r=1$ and $r=2$ cases) and computing cost simultaneously. 

%\begin{figure}[bt]
%    \begin{center}
%        \addpgfplot[0.5\linewidth]{pareto_tradeoff_n400.tikz}
%    \end{center}
%    \caption{Expected latency $\E{T}$ versus the expected cost $\E{\Cost}$ for $\PPareto{2}{2}$ and $n=400$, by
%    varying $p$ along each curve in the range of $[0, 1]$. For small $p$, we can reduce both latency and cost
%    simultaneously.
%    }
%    \label{fig:pareto_ET_vs_ECloud}
%\end{figure}

%With \Cref{theorem:single_fork_pareto}, we can compute the latency and cost of both cases of keeping or killing
%the original copy of a straggling task, and compare to see which strategy is . 

%identify the values of $p$ for the relaunching $(l=0)$ case when
%the corresponding single-fork policy is sub-optimal in both $\E{T}$ and $\E{C}$.

%In \Cref{lem:pareto_suboptimal_policies} we 
%identify the values of $p$ for the relaunching $(l=0)$ case when
%the corresponding single-fork policy is sub-optimal in both $\E{T}$ and $\E{C}$.

%For example for the $r=1$ and relaunch ($l=0$) case, we can solve \Cref{eq:pareto_sub_optimal} to show that all
%policies with $p < p_1^* \approx 0.05$ are sub-optimal, where $p_1^*$ is marked in
%\Cref{fig:pareto_ET_vs_ECloud}. Similarly, for cases $r=0$ and $r=2$, the sub-optimal ranges $[0, p_0^*]$ and $[0, p_2^*]$ are shown respectively in \Cref{fig:pareto_ET_vs_ECloud}.

%We conjecture that the convex hull of the curves for different $r$ and $l$ gives the optimal latency-cost
%trade-off. Points on the hull, which lie between some two curves can be achieved by time-sharing between the
%corresponding two policies. 

\section{Empirical execution time distributions}
\label{sec:heuristic}
%\TODO{Find a better section title}

In practice, it may be difficult to fit the empirical behavior of the task execution time to a
well-characterized distribution, thus making the latency-cost analysis using the framework presented in
\Cref{sec:single_fork} difficult. In this section we propose an algorithm to estimate the latency and cost from
the empirical distribution of task execution time.
This enables users to evaluate the latency-cost trade-off of various replication strategy using execution trace
directly, instead of a fitted execution time distribution. 
%To overcome this, we propose using the bootstrapping
%method~\cite{efron1986bootstrap} in \Cref{subsec:est_perf_metrics}, where we essentially treat the empirical
%distribution of execution time as the true distribution and analyze the system performance in terms of latency
%and cost via using random sampling methods. 
Applying our algorithm to the Google Cluster Trace data
\cite{web_google_cluster_data_doc}, we show that it is possible to 
improve upon the performance of the default replication policy in MapReduce-style frameworks.
%obtain improved scheduling performance in real world scenarios as well.

\subsection{Latency and Cost Estimation}
\label{subsec:est_perf_metrics}
To estimate the latency and cost from empirical execution time samples, we apply the bootstrapping method~\cite{efron1986bootstrap} that uses the empirical distribution as an approximation of the true distribution. 

%The key idea in \emph{bootstrapping} is to use the empirical distribution as an approximation of the sampling
%distribution, and by drawing samples from this approximated sampling distribution, we can often estimate a
%statistic of the sampling distribution quite well, in the sense that the estimated statistic converges to the
%true sample statistic. 
%For example, to estimate the median of the execution time distribution $X$, we may sample
%$n$ times from the empirical distribution of execution time $\hF_X(\cdot)$, and calculate the median of the $n$
%samples $\hx_1, \hx_2, \ldots, \hx_n$. 

Since the performance metrics $\E{T}$ and $\E{C}$ are functions of both $X$ and $Y$, we need samples for both $X$ and $Y$. Drawing samples of $Y$ is more involved, especially for the case of killing the original task. To handle this, we compute $\hF_Y(\cdot)$ using \eqref{eqn:Y_defn}, thus avoiding excessive sampling. We present the algorithm for performance characterization in \Cref{algo:est_perf_metrics}.

\begin{algorithm}[t]
\begin{algorithmic}
	\State \textbf{INPUT}: $\bx = [x_1, x_2, \ldots, x_n]$, $n$ task execution duration samples (no replication, no original task killing)
	\State Compute the empirical \cdfunc\ $\hF_X(x)$ from $\bx$ 
	\State Compute \cdfunc\ $\hF_Y(y)$ using \eqref{eqn:Y_defn}
	\For{ $i = 1, 2, \ldots m$}
		\State Draw $n$ samples $\bhx = [\hx_1, \hx_2, \ldots, \hx_n]$ from $\hF_X$
		\State Sort $\bhx$ in ascending order: $[\hx_{(1)}, \hx_{(2)}, \ldots, \hx_{(n)}]$
		\State $k \gets n(1-p)$; $k' \gets np$ 
		\State $\tilde{T}_1^{(i)} \gets \hx_{(k)}$ (the $k$-th smallest sample in $\bhx$)
		\State $\tilde{C}_1^{(i)} \gets \sum_{j=1}^k \hx_{(j)}$ 
		\State Draw $k'$ samples $\bhy = [\hy_1, \hy_2, \ldots, \hy_{k'}]$ from $\hF_Y$ 
		\State $\tT_2^{(i)} \gets \max_{1 \leq j \leq k'} \hy_j$ 
		\State $Y_{sum}^{(i)} \gets \sum_{j=1}^{k'} \hy_j$ 
		\State $\tC_2^{(i)} \gets pn\tilde{T}_1^{(i)} + (r+1)Y_{sum}^{(i)}$
		\State $\tilde{T}^{(i)} \gets  \tilde{T}_1^{(i)} + \tT_2^{(i)}$
		\State $\tilde{C}^{(i)} \gets  \frac{1}{n}\left[ \tilde{C}_1^{(i)} + \tC_2^{(i)} \right]$
    	\EndFor
    \State $\tilde{T} \gets$ mean of $\tilde{T}^{(i)}$ for $i = 1, 2, \ldots m$
    \State $\tilde{C} \gets$ mean of $\tilde{C}^{(i)}$ for $i = 1, 2, \ldots m$
	\State \textbf{OUTPUT}: $[\tT, \tC]$
\end{algorithmic}
\caption{Latency and cost estimation}
\label{algo:est_perf_metrics}
\end{algorithm}

By \Cref{thm:central_order_stats}, the standard deviation of the error in estimating $\E{C}$ and $\tT_1$, first term in $\E{T}$, converges to zero as $O(1/\sqrt{m n})$, where $m$ is the number of times the sampling procedure
is repeated. And generally $\tT_2$, the maximum order statistic term in $\E{T}$, converges to zero as $\BigO{1/\sqrt{m}}$. Thus, the estimation of $\tilde{C}$ is more robust than that of $\tilde{T}$.  Nonetheless, with
large enough $m$, we can make the estimation errors of both metrics small enough. 

\subsection{Demonstration using Google Cluster Trace}
The Google Cluster Trace data \cite{web_google_cluster_data_doc} gives timestamps of events such as SCHEDULE,
EVICT, FINISH, FAIL, KILL etc.\ for each of the tasks of computing jobs that are run on Google's cluster
machines. In this section we apply \Cref{algo:est_perf_metrics} to two jobs in the Google Cluster Trace, and
study the latency-cost trade-offs for these real-world task service distributions. 

In our demonstration 
%we use version 2.1 of the Google Cluster Trace data, which is first released on February, 2011. 
we only consider tasks with SCHEDULE and FINISH times, as we would like to obtain samples that represent a
normal execution (not killed or evicted). In a few rare cases, a task is associated with multiple SCHEDULE and
FINISH events due to duplicate execution. For these we choose to keep the first occurrences in each event category. 

\begin{figure}[t!b]
    \begin{subfigure}[b]{0.32\textwidth}
        \centering
        \includegraphics[width=\linewidth]{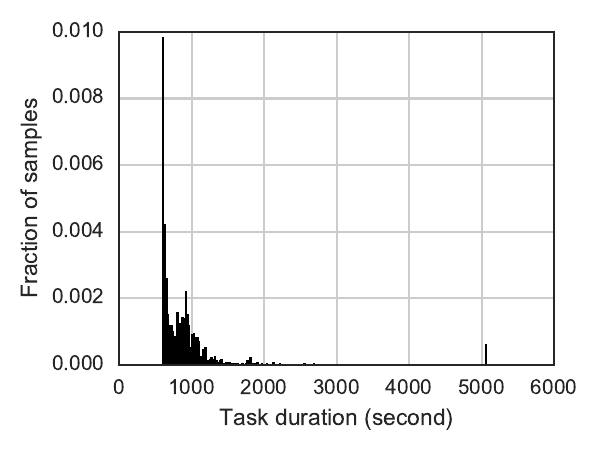}
        \captionsetup{format=hang}
        \caption{Job 1: Google cluster Job 6252284914 (1026 tasks)}
        \label{fig:google_trace_hist}
    \end{subfigure}%
    \hspace*{\fill}%
    \begin{subfigure}[b]{0.32\textwidth}
        \centering
        \includegraphics[width=\linewidth]{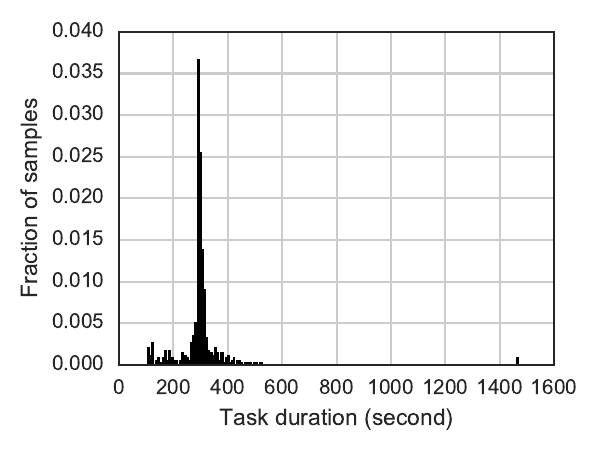}
        \captionsetup{format=hang}
        \caption{Job 2: Google cluster Job 6252315810 (488 tasks)}
        \label{fig:google_trace_hist_2}
    \end{subfigure}
    \hspace*{\fill}%
    \begin{subfigure}[b]{0.32\textwidth}
        \centering
        \includegraphics[width=\linewidth]{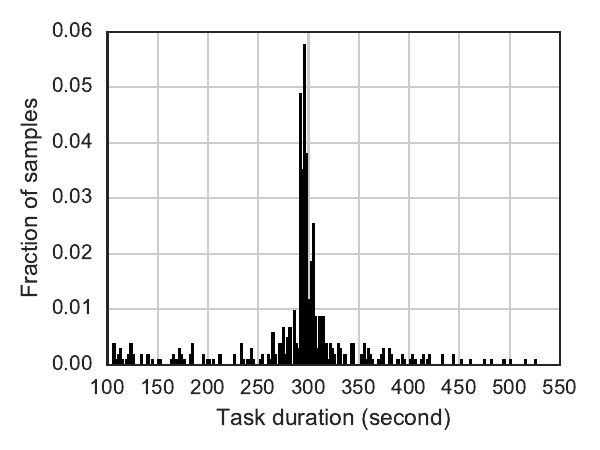}
        \captionsetup{format=hang}
        \caption{Job 3: Tail-shortened trace of Job 6252315810 (485 tasks)}
        \label{fig:google_trace_hist_2_mod}
    \end{subfigure}
    \caption{Normalized histogram of the task execution times} 
\end{figure}

\begin{figure}[t]
    \hspace*{\fill}%
    \begin{subfigure}[b]{0.45\textwidth}
        \centering
        \includegraphics[width=3.2in]{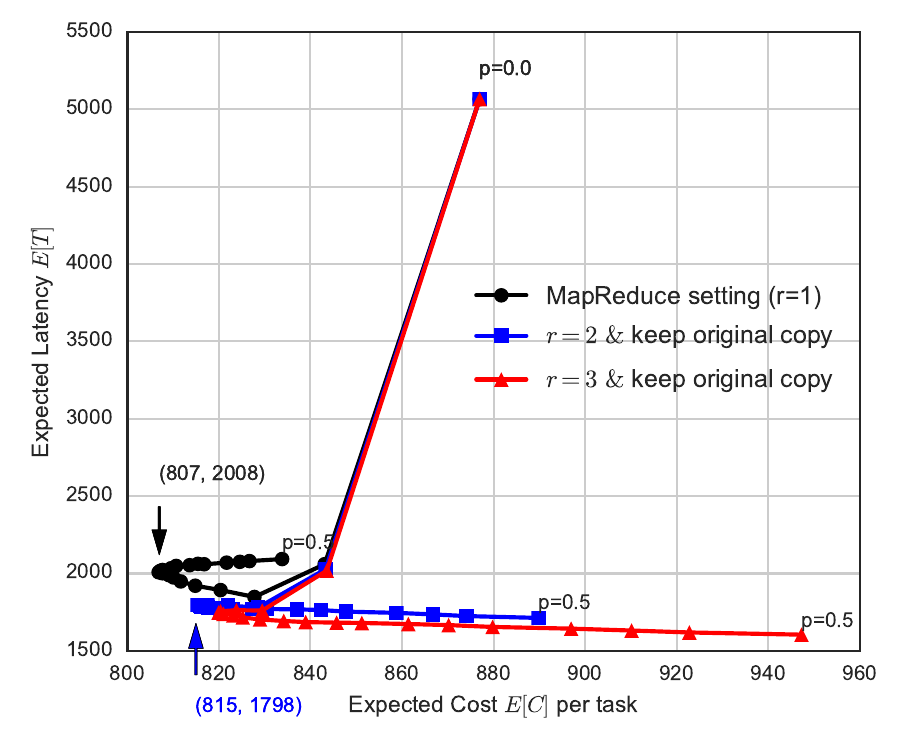}
        \caption{Trade-off with original copy kept $\SingleForkKeepText$}
    \end{subfigure}
    \hspace*{\fill}%
    \begin{subfigure}[b]{0.45\textwidth}
        \centering
        \includegraphics[width=3.2in]{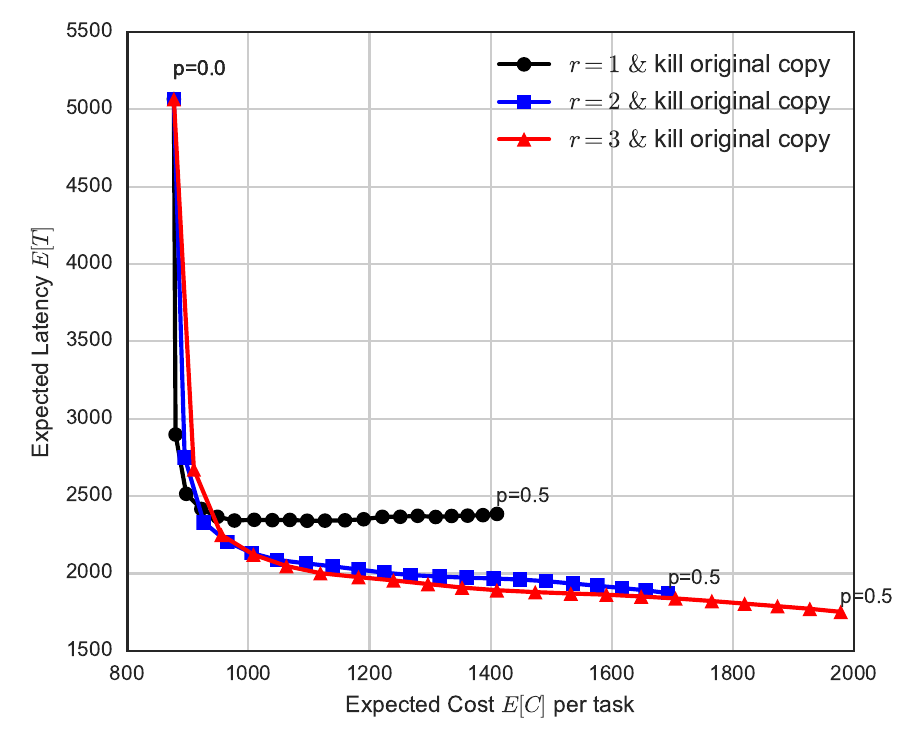}
        \caption{Trade-off with original copy killed $\SingleForkKillText$}
    \end{subfigure}
    \hspace*{\fill}%
    \caption{
    The $\E{T}$-$\E{C}$ trade-off for Job 1 (ID 6252284914) with $1026$ tasks. 
    Each pair of adjacent dots corresponds to change in $p$ by $0.01$.
    }
    \label{fig:heuristic_result}
\end{figure}

\begin{figure}[t]
    \hspace*{\fill}%
    \begin{subfigure}[b]{0.45\textwidth}
        \centering
        \includegraphics[width=3.2in]{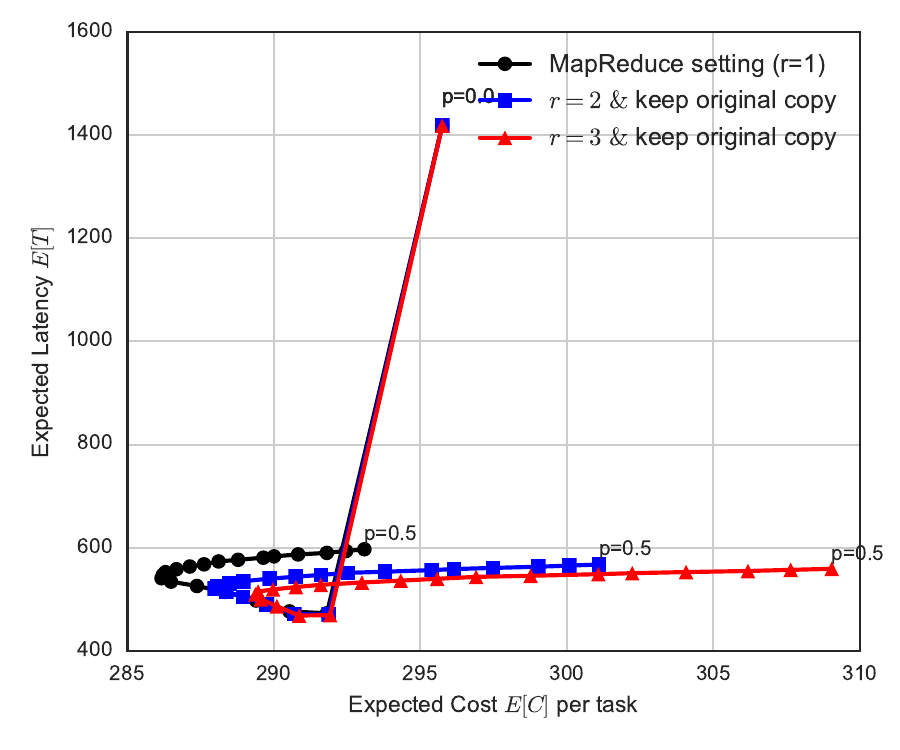}
        \caption{Trade-off with original copy kept $\SingleForkKeepText$}
    \end{subfigure}
    \hspace*{\fill}%
    \begin{subfigure}[b]{0.45\textwidth}
        \centering
        \includegraphics[width=3.2in]{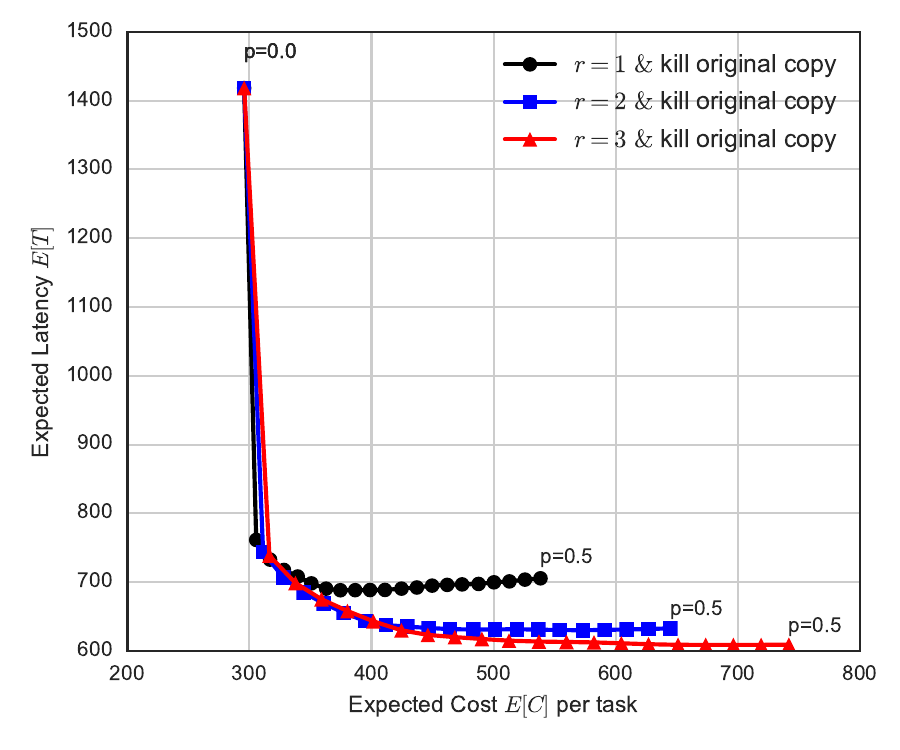}
        \caption{Trade-off with original copy killed $\SingleForkKillText$}
    \end{subfigure}
    \hspace*{\fill}%
    \caption{
    The $\E{T}$-$\E{C}$ trade-off for Job 2 (ID 6252315810) with $488$ tasks.
    Each pair of adjacent dots corresponds to change in $p$ by $0.01$.
    }
    \label{fig:heuristic_result_2}
\end{figure}

\begin{figure}[t]
    \hspace*{\fill}%
    \begin{subfigure}[b]{0.45\textwidth}
        \centering
        \includegraphics[width=3.2in]{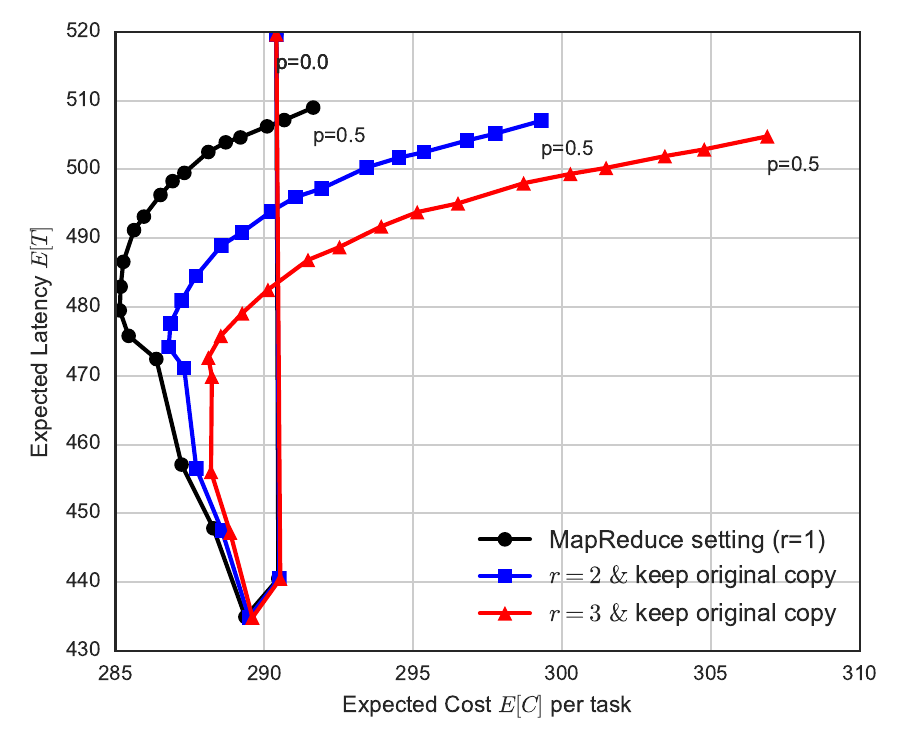}
        \caption{Trade-off with original copy kept $\SingleForkKeepText$}
        \label{fig:heuristic_result_2_mod_keep}
    \end{subfigure}
    \hspace*{\fill}%
    \begin{subfigure}[b]{0.45\textwidth}
        \centering
        \includegraphics[width=3.2in]{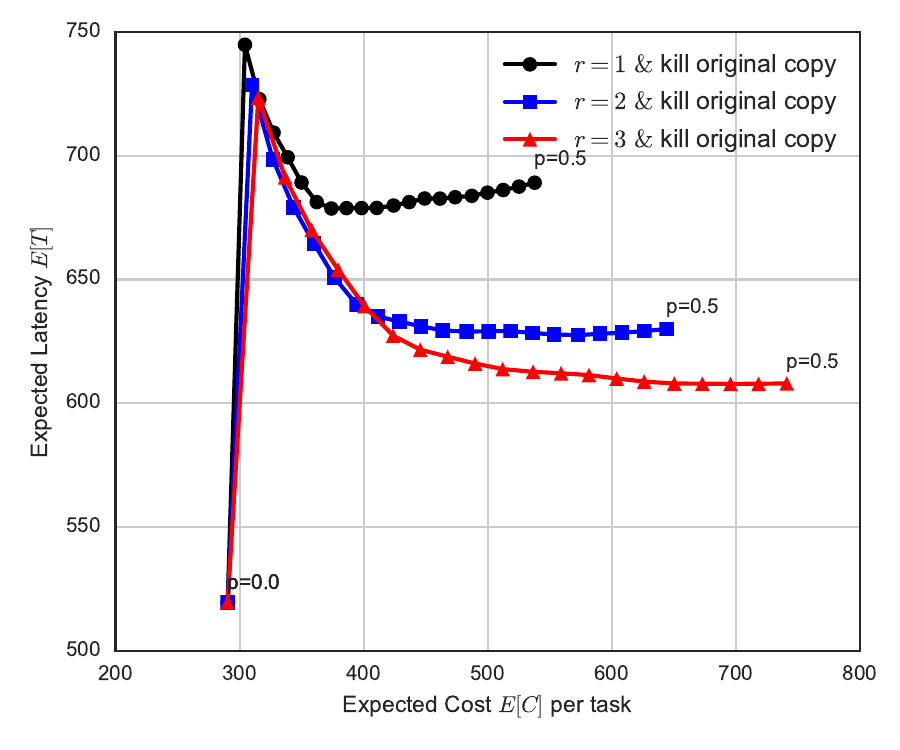}
        \caption{Trade-off with original copy killed $\SingleForkKillText$}
    \end{subfigure}
    \hspace*{\fill}%
    \caption{
    The $\E{T}$-$\E{C}$ trade-off for the Job 3 (tail-shortened Job 2) with $485$ tasks.
    Each pair of adjacent dots corresponds to change in $p$ by $0.01$.
    }
    \label{fig:heuristic_result_2_mod}
\end{figure}

We choose two jobs (Job ID 6252284914 and 6252315810) with different numbers of tasks. For each task in a job, we obtain the task execution time by calculating the time difference between SCHEDULE and FINISH. The normalized histograms of the task execution times of the two jobs are shown in \Cref{fig:google_trace_hist} and \Cref{fig:google_trace_hist_2} respectively. Both the distributions have straggling tasks whose execution time is significantly longer than average. To emphasize the importance of such stragglers, we modify the trace for Job 6252315810 by removing the 3 samples with execution time longer than 1400 seconds, leading to the execution time distribution shown in \Cref{fig:google_trace_hist_2_mod}.
We then apply these execution time samples as inputs to \Cref{algo:est_perf_metrics} with $m=1000$. By varying the value of $r$ ($r \in
\Set{1,2,3}$) and $p$ ($0 \leq p \leq 0.5$), we plot the $\E{T}$-$\E{C}$ trade-offs for all three jobs in \Cref{fig:heuristic_result,fig:heuristic_result_2,fig:heuristic_result_2_mod}.


For the two Google cluster jobs (Job 1 and 2), we observe that a small amount of replication (small $p$) reduces both $\E{T}$ and $\E{C}$ significantly, demonstrating the effectiveness of replication for real-world 
execution time distributions. In both cases, it is better to replicate while \emph{keeping} the original task, because at the ``fork'' point, the additional time needed for the original copy to finish is more likely to be shorter than the execution time of a new copy. We also observe that for the Job 2 (Job ID 6252315810), too much redundancy may hurt, because at some point
increasing $p$ actually leads to increases in both $\E{T}$ and $\E{C}$. However, this phenomenon does not exist for Job 1 (Job ID 6252284914) when $r=2$ or $r=3$. We conjecture this is due to the tail in \Cref{fig:google_trace_hist} is heavier than that in \Cref{fig:google_trace_hist_2}.

We recall that the back-up tasks option in MapReduce uses $r=1$ and keeps the original task, and show that for certain jobs it may be more desirable to improve the performance trade-off by using more replicas, such as in Job 1, where a higher $r$ could lead to lower latency $\E{T}$ with a slightly higher cost $\E{C}$.
For example, $\SingleForkKeep{p, r=1}$ achieves $(\E{C}, \E{T}) = (807, 2008)$, while $\SingleForkKeep{p, r=2}$ achieves $(\E{C}, \E{T}) = (815, 1798)$.
For Job 2, the trade-off improvement via using a higher $r$ is less significant, as 
\Cref{fig:heuristic_result_2} indicates. Finally, for both jobs we observe that increasing $r$ has a diminishing effect on the reduction of $\E{T}$.

For the tail-shortened trace histogram in \Cref{fig:google_trace_hist_2_mod}, killing the original copy
increases the latency, because it is too ``impatient''---the original copy is likely to finish before a new copy
of the task. On the other hand, if we keep the original copy, adding a small amount of redundancy can
reduce latency and computing cost simultaneously, as shown in \Cref{fig:heuristic_result_2_mod_keep}.  Lastly,
\Cref{fig:heuristic_result_2_mod} indicates that killing and replicating tasks can lead to a worse performance
trade-off, so one needs to apply replication with care.

%The above examples demonstrate that replication often reduces $\E{T}$ and $\E{C}$ simultaneously. A small amount of replication usually suffices and too much replication leads to a sharp increase in $\E{C}$.

%To summarize, once we are able to characterize the trade-off between latency $\E{T}$ and
%computing cost $\E{C}$, we can choose the suitable operating point and the corresponding scheduling policy. This
%sometimes could leads to lower latency and lower cost simultaneously and could be more desirable than the
%MapReduce setting for certain execution time distributions.

\subsection{Scheduling policy selection}
With the trade-off between latency $\E{T}$ and computing cost $\E{C}$ provided in \Cref{algo:est_perf_metrics},
a user can formulate an optimization problem to choose the best scheduling policy based on one's sensitivity to
latency and computing cost. In addition, one can incorporate additional constraints, such as $r_{\max}$, the
maximum number of copies to replicate, due to the communication overhead of issuing and canceling tasks.

For example, a latency-sensitive user may choose to define the optimal scheduling policy via the following constrained
optimization problem:
\begin{alignat}{2}
 & \text{minimize} &\quad& \E{T(\pi)},
 \label{eq:latency_sensitive_opt}
 \\
 & \text{subject to} &\quad& \E{C(\pi)} \leq \E{C(\pi_0)}, \nonumber \\
 &  &\quad& r \leq r_{\max},
 \nn
\end{alignat}
where $\pi_0$ is the baseline scheduling policy without replication and $r_{\max}$ the maximum allowed number of copies for a task. 
%Therefore, the optimal scheduling policy $\pi^*$ satisfies
%\begin{align}
%\pi^* = \argmin_{\substack{\pi \in \Set{\SingleForkKeepText, \SingleForkKillText}\;\setst\;\E{C(\pi)} \leq \E{C(\pi_0)}, \\ p \in [0, 1],\; r \leq r_{\max}}} 
%            \E{T(\pi(p, r))}
%\label{eq:latency_sensitive_opt}
%\end{align}
On the other hand, a cost-sensitive user may choose to define the optimal scheduling policy via the following optimization problem:
\begin{alignat}{2}
 \label{eq:cost_sensitive_opt}
 & \text{minimize} &\quad& \E{T(\pi)} + \lambda n \E{C(\pi)},
 \\
 & \text{subject to} &\quad& r \leq r_{\max},
 \nn
\end{alignat}
where $\lambda$ indicates the relative importance of computing cost, because $\E{C}$ is approximately proportional to the cost of cloud computing instances. While it is difficult to determine closed-form optimal solutions to \eqref{eq:latency_sensitive_opt} and \eqref{eq:cost_sensitive_opt}, we observe that constrained optimization methods such as the Constrained Optimization BY Linear Approximation (COBYLA) method \cite{powell_cobyla_2007} are effective in searching for the optimal solution due to the low dimensionality of the search space. In \Cref{tab:opt_policies}, we present the scheduling policies obtained via these two different optimization formulations.
%Therefore, the optimal scheduling policy $\pi^*$ satisfies
%\begin{align}
%\pi^* = \argmin_{\substack{\pi \in \Set{\SingleForkKeepText, \SingleForkKillText}, \\ p \in [0, 1],\; r \leq r_{\max}}} 
%            \E{T(\pi(p,r))} + \lambda n \E{C(\pi(p,r))},
%\label{eq:cost_sensitive_opt}
%\end{align}

\begin{table}[h]
\footnotesize
\centering
\begin{tabular}[b]{|c|cc|ccccc|ccccc|}
\hline
& \multicolumn{2}{c|}{\textbf{Baseline}}
& \multicolumn{5}{c|}{\textbf{Latency-sensitive}}
& \multicolumn{5}{c|}{\textbf{Cost-sensitive with $\lambda=0.1$}}
\\
\hline
\textbf{Job}
& $\E{T}$ & $\E{C}$
& $p^*$ & $r^*$ & keep/kill & $\E{T}$ & $\E{C}$
& $p^*$ & $r^*$ & keep/kill & $\E{T}$ & $\E{C}$
\\
\hline
\textbf{Job 1}
& 5068 & 882
& 0.343 & 4 & keep & 1676 & 881
& 0.234 & 1 & keep & 2213 & 806
\\
\hline
\textbf{Job 2}
& 1418 & 296
& 0.038 & 4 & keep & 463 & 291
& 0.181 & 4 & keep & 542 & 286
\\
\hline
\textbf{Job 3}
& 520 & 290
& 0.044 & 4 & keep & 432 & 290  
& 0.173 & 1 & keep & 480 & 285
\\
\hline
\end{tabular}
\caption{Scheduling policy obtained via latency-sensitive optimization in \eqref{eq:latency_sensitive_opt} and cost-sensitive optimization in \eqref{eq:cost_sensitive_opt}.} 
\label{tab:opt_policies}
\end{table}

\section{Concluding remarks}
\label{sec:conc_remarks}

\subsection{Main Implications}
Replication of the slowest tasks of a computing job (straggling tasks) has been observed to be highly effective
in practice to speed-up job completion. In this paper we provide a theoretical framework to understand the
effect of straggler replication on the job completion latency, and the additional computing time spent on
running the replicas. Our latency-cost analysis gives the insight that the scaling of job completion latency
with the number of tasks depends on the tail of the per-task execution time. We identify regimes where
replicating a small fraction of stragglers can drastically reduce latency and computing cost simultaneously. 
With the guidance from this asymptotic analysis, we propose a bootstrapping-based algorithm to estimate the
latency and cost from empirical traces of execution time. The effectiveness of this algorithm is demonstrated on
the Google Cluster Trace data, where we show that careful choice of the replication strategy can improve the
latency-cost trade-off as compared to the default option in MapReduce.

%Using tools from extreme value theory, we characterize the latency-cost trade-off in terms of the task execution time distribution $F_X$. We focus on three parameters of a replication strategy: 1) fraction of slowest tasks of a job that are considered as stragglers, 2) number of replicas of each straggling task, and 3) whether we should kill the original copy of the task and relaunch it on a new machine. 

\subsection{Future Directions}
Generalizations of this straggler replication model include considering heterogeneous servers, dependencies
between tasks (some tasks need to complete in order to begin others), and taking into account queueing delay of
tasks as considered in
\cite{gauri_allerton_2015,gauri_thesis_2016,gardner_sigmetrics_2015,gardner_redundancy_d_2016} for the single
task case. Another direction is to analyze approximate computing, where we need only a subset of the tasks of a
job to complete, a relevant model for information retrieval and machine learning jobs. This idea is developed in
the context of coded distributed storage in \cite{gauri_yanpei_emina_jsac,mds_queue}. We also aim to develop an
algorithm that learns the task execution time distribution $F_X$ online, and use it to decide when and how many
replicas to launch. This has an exploration-exploitation trade-off, similar to the multi-arm bandit problems
studied in reinforcement learning~\cite{sutton_introduction_1998}. 

More broadly, our analysis framework can be applied to other systems with stochastically varying components, for
example, in crowdsourcing, each worker may take a variable amount of time to complete a
task~\cite{wang_thesis_2014}.

\appendix
\section{Appendix}
\label{sec:appendix}
\subsection{Results from Order Statistics}
\label{sec:central_order_stats}
%For an order statistic $\OSn{k}$, we called it a {central order statistic} if $k \approx np$ for some $p \in (0,1)$.  In this case, $\OSn{k}$ is asymptotically normal, concentrated around the $p$-th quantile of $X$, as indicated by the following result called the Central Value Theorem (Theorem 10.3 in \cite{david_order_2003}). 
%
\begin{thm}[Central Value Theorem (Theorem 10.3 in \cite{david_order_2003})]
\label{thm:central_order_stats}
Given $X_1, X_2, \allowbreak \ldots, \allowbreak X_n \beiid F_X$, if $0 < p < 1$ and $0 < f(x_p) < \infty$, where $x_p = F_X^{-1}(p)$, then
for $k = np + \SmallO{\sqrt{n}}$, the $k^{th}$ order statistic is asymptotically normal,
\begin{equation*}
    \OSn{k}
    \convergeinprob
    \PGaussian{x_p}{\frac{p(1-p)}{n f^2(x_p)}}
\end{equation*}
where $f(\cdot)$ is the \pdfunc\ corresponds to $F_X$ and $\convergeinprob$ denotes convergence in probability as $n \rightarrow \infty$.
\end{thm}
%Therefore, when $n$ is large, $\OSn{k}$ is tightly concentrated around $x_p$. 
%
%\section{Extreme order statistics}
%\label{sec:extreme_order_stats}
{Extreme value theory} (EVT) is an asymptotic theory of extremes, \ie, minima and maxima. It shows that if a distribution belongs to one of three families of distributions \Cref{thm:domain_of_attractions}), then its maxima can be well characterized asymptotically as given by \Cref{thm:ev_thm}, which is also referred to as the
Fisher-Tippett-Gnedenko Theorem (Theorem 1.1.3 in~\cite{haan_extreme_2006}).
\begin{thm}[Domains of attraction]
    \label{thm:domain_of_attractions}
    A distribution function $F_X$ has one of the following domains of attraction if it satisfies the conditions of the extreme value
    distribution $G(x)$ if and only if
    \begin{enumerate}
        \item $F_X \in \DAG$ if and only if there exists $\eta(x) > 0$ such that
            \begin{equation*}
                \lim_{x \rightarrow \CDUpper{F}^-} \frac{\brF(x + t \eta(x))}{\brF(x)} = e^{-t};
            \end{equation*}
        \item $F_X \in \DAF$ if and only if $\CDUpper{F} = \infty$ and
            \begin{equation*}
                \lim_{\ntoinf[x]} \frac{\brF(tx)}{\brF(x)} = t^{-\xi}, \quad t > 0;
            \end{equation*}
        \item $F_X \in \DAW$ if and only if $\CDUpper{F} < \infty$ and
            \begin{equation*}
                \lim_{x \rightarrow 0^+} \frac{\brF(\CDUpper{F} - tx)}{\brF(\CDUpper{F} - x)} = t^{\xi}, \quad t > 0;
            \end{equation*}
    \end{enumerate}
    where $\CDUpper{x} = \sup \{ x : F_X(x) < 1 \}$, the upper end point of the distribution $F_X$.
\end{thm}

Intuitively, 
$F \in \DAG$ corresponds to the case that $\brF$ has an exponentially decaying tail, 
$F \in \DAF$ corresponds to the case that $\brF$ has heavy tail (such as 
polynomially decaying), 
and 
$F \in \DAW$ corresponds to the case that $\brF$ has a short tail with finite
upper bound.

\begin{thm}[Extreme Value Theorem]
    \label{thm:ev_thm}
    Given $X_1$, $\ldots, X_n \allowbreak \beiid F$, if there exist sequences of constants
    $a_n>0 $ and $b_n\in \reals$ such that
    \begin{align}
        \label{eqn:ev_dist}
        \Prob{(\OSn{n}-b_n)/a_n \leq x} \rightarrow G(x)
    \end{align}
    as $n \rightarrow \infty$ and $G(\cdot)$ is a non-degenerate distribution. 
The extreme value distribution $G(x)$ and the values of  $a_n$ and $b_n$ depend on the domain of attraction (and
hence the tail behavior) of $F_X$ given by \Cref{thm:domain_of_attractions}.
\begin{enumerate}
    \item For $F_X \in \DAG$,
        \begin{align}
            a_n &= \eta\left( F^{-1}(1-1/n) \right), \label{eqn:a_n_gumbel}\\
            b_n &= F^{-1}(1-1/n) \label{eqn:b_n_gumbel}\\
            G(x)& = \GumbelDist(x) = \exp\left\{-\exp\left(-x\right)\right\} \label{eqn:gumbel_law}
        \end{align}
        where $\GumbelDist(x)$ is called the {Gumbel distribution}.
    \item For $F_X \in \DAF$,
        \begin{align}
            a_n &= F^{-1}(1-1/n), \label{eqn:a_n_frechet} \\
            b_n &= 0, \label{eqn:b_n_frechet} \\
            G(x)  &= \FrechetDist(x) =\begin{cases} 
                0 & x \leq 0 \\ 
                \exp\left\{-x^{-\xi}\right\} & x > 0
            \end{cases}. \label{eqn:frechet_law}
        \end{align}
        where $\FrechetDist(x)$  is called the {\Frechet\ distribution}.
    \item For $F_X \in \DAW$,
        \begin{align}
            a_n &= \CDUpper{F} - F^{-1}(1-1/n),\\
            b_n &= \CDUpper{F}, \\
            G(x) &= \RWeibullDist(x) =\begin{cases} 
                \exp\left\{-\left( - x \right)^\xi\right\} & x<0, \\ 
                1 & x\geq 0.
            \end{cases}  \label{eqn:weibull_law}
        \end{align}
        where $\RWeibullDist(x)$ is called the {reversed-Weibull distribution}.
\end{enumerate}
\end{thm}
Based on \Cref{thm:ev_thm}, we can derive the expected value of extreme values, as shown in
\Cref{lem:E_evt_dists}.
\begin{lem}[Expected Extreme Values]
    \label{lem:E_evt_dists}
    \begin{align*}
        \E{\GumbelDist} &= \EMConstantSymbol,
        \\
        \E{\FrechetDist} &= \begin{cases}
            \Gamma\left( 1 - 1/\xi \right)
            & \xi > 1 \\
            +\infty & \mathrm{otherwise},
        \end{cases} \\
        \E{\RWeibullDist} &= -\Gamma\left( 1 + 1/\xi \right),
    \end{align*}
    where $\EMConstantSymbol$ is the \EulerConstant\ and 
    $\Gamma(\cdot)$ is the Gamma function, \ie,
    \begin{equation*}
        \Gamma(t) \defeq \int_0^{\infty} x^{t-1} e^{-x} \, dx
        .
    \end{equation*}
\end{lem}

We can also characterize the limit distribution of the sample extreme $\OSn{1}$
analogously via \Cref{thm:ev_thm} by 
\begin{equation*}
    X_{1:n} = \min\Set{X_1, \ldots, X_n} = - \max\Set{-X_1, \ldots, -X_n}.
\end{equation*}
It is worth noting that the distribution function for $-X$ may be in a different
domain of attraction from that of $X$.

%\section{Proofs of single-fork analysis}
%label{sec:proof_single_fork}
%In this appendix we give proofs of the results  in \Cref{sec:single_fork}.
%

\subsection{Proofs of Single Fork Analysis}
\label{sec:proof_single_fork_general}

\begin{proof}[of \Cref{thm:single_fork_gen}]
The expected latency $\E{T}$ can be divided into two parts: before and after replication. % of $pn$ straggling tasks. 
\begin{align}
\E{T} &= \E{\Ta} + \E{\Tb}, \nonumber \\
&= \E{X_{(1-p)n:n} } + \E{ \max_{j=1, 2, \dots, pn} Y_j}, \nonumber \\
&= F_X^{-1}(1-p) + \E{Y_{pn:pn}}. \label{eqn:latency_1}
\end{align}
The time before forking $\Ta$ is the time until $(1-p)n$ of the $n$ tasks launched at time $0$ finish. Thus, its expected value $\E{\Ta}$ is the expectation of the $(1-p)n ^{th}$ order statistic $X_{(1-p)n:n}$ of $n$ i.i.d.\ random variables with distribution $F_X$. By the Central Value Theorem stated as \Cref{thm:central_order_stats}, for $n \rightarrow \infty$, this term converges to inverse CDF value $F_X^{-1}(1-p)$.

At this forking point, the scheduler introduces replicas of the $pn$ straggling tasks. The distribution $F_Y$ of the residual execution time (minimum over the $r+1$ replicas). First consider $\SingleForkKillText$ where the original copy is killed. The residual execution
time distribution $F_Y$ (after time $T^{(1)}$ when the replicas are added) of each task is the minimum of $r+1$ i.i.d.\ random variables with distribution $F_X$. Hence,
\begin{align}
\Pr(Y>y) &= \Pr( \min (X_1, X_2, \dots X_{r+1} ) > y ) ,\\
\fccdf[Y]{y} &= \fccdf[X]{y}^{r+1} \quad \text{ for } \SingleForkKillText.
\end{align}
For $\SingleForkKeepText$, there is $1$ original replica and $r$ new replicas of each of the straggling tasks. Thus, the tail distribution $\fccdf[Y]{y} = 1- F_Y(y)$ is given by
\begin{align}
\Pr(Y>y) &= \Pr( X_1 > y + T^{(1)} \vert X_1 > T^{(1)}) \cdot \Pr( \min (X_2,\dots X_{r+1} ) > y ), \\
%\intertext{\ie, when $y\geq0$,}
\fccdf[Y]{y} &= \frac{\fccdf[X]{y +T^{(1)} }}{ \fccdf[X]{T^{(1)}}} \fccdf[X]{y}^{r}.
\end{align}
As the number of tasks $n \rightarrow \infty$ by \Cref{thm:central_order_stats} we have $T^{(1)} \rightarrow F_X^{-1}(1-p)$. Hence,
\begin{align}
\fccdf[Y]{y} &= \frac{\fccdf[X]{y +F_X^{-1}(1-p) }}{p} \fccdf[X]{y}^{r} \quad \text{ for } \SingleForkKeepText.
\end{align}
The second term $\E{\Tb}$ in \eqref{eqn:latency_1} is the expected value of the maximum of $pn$ i.i.d.\ random variables with distribution $F_Y$.

Recall from \Cref{defn:cost} that the expected cost $\E{C}$ is the sum of the running times of all machines, normalized by the number of tasks $n$. We can analyze $\E{C}$ by dividing it into sum of machine runtimes before and after forking.
\begin{align}
    \E{\Cost} &= \E{\Ca} + \E{\Cb} \label{eqn:cost_1} , \\
%\end{align}
%where $\E{\Ca}$ and $\E{\Cb}$ can be evaluated as follows.
%\begin{align}
  \E{\Ca} &= \frac{1}{n} \sum_{i=1}^{(1-p) n} \E{\OSn{i}} + \frac{n p}{n} \E{ \Ta} \label{eqn:cost_2},\\
   &= \frac{1}{n} \sum_{i=1}^{(1-p) n} F_X^{-1} \left(\frac{i}{n} \right) + p F_X^{-1}(1-p) \label{eqn:cost_3},\\
   &= \int_{0}^{1-p} F_X^{-1}(h) dh + p F_X^{-1}(1-p) \label{eqn:cost_4}.
\end{align}

\begin{align}
 \E{\Cb} 
  &= \frac{1}{n}  \sum_{j=1}^{pn} (r+1) \E{Y_j} \label{eqn:cost_5},\\
  &= (r+1)  p \cdot \E{Y} . \label{eqn:cost_6}
\end{align}
The cost before forking $\E{\Ca}$ consists of the cost for the $(1-p)n$ tasks that finish first, plus the cost
for the $pn$ straggling tasks. The first term in \eqref{eqn:cost_2} is the sum of the expected values of the
smallest $(1-p)n$ execution times. Using \Cref{thm:central_order_stats}, we can show that the $i^{th}$ term in
the summation converges to $F_X^{-1}(i/n)$ as $n \rightarrow \infty$. Expressing the sum as an integral over $h
= i/n$ we get the first term in \eqref{eqn:cost_4}. The second term in \eqref{eqn:cost_2}, is the normalized
running time of the $pn$ straggling tasks before forking. Substituting $\E{\Ta}$ from \eqref{eqn:latency_1} and
simplifying, we get \eqref{eqn:cost_4}. 

The cost after forking, $\E{\Cb}$ is the normalized sum of the runtimes of the $r+1$ replicas of each of the
$pn$ straggling tasks. The residual execution time of the $j^{th}$ straggling task is $Y_j \sim F_Y$. Since the
scheduler kills all replicas as soon as one replica finishes, the expected runtime for the $j^{th}$ straggling
task is $(r+1) \E{Y_j}$. Thus, the cost in \eqref{eqn:cost_5} is the sum of $(r+1) \E{Y_j}$ over the $pn$ tasks,
normalized by $n$. Since $Y_j$ are i.i.d, we can reduce this to \eqref{eqn:cost_6}.
\end{proof}

%To prove \Cref{lem:stage2_latency}, we characterize the expected maximum of a large number of random variables
%using \Cref{thm:ev_thm}. 
%%% % % % % % % % % % % % % % % % % % % % % % % % % % % % % % % % % % % % % % %
%\begin{proof}[of \Cref{lem:stage2_latency}]
%    We can use \Cref{lem:DA_FY} to find the domain of attraction of $F_Y$. Then from \eqref{eqn:ev_dist} we have
%    \begin{align*}
%        \E{\OSn[Y]{n}} = \tla_n \E{G(y)} + \tlb_n,
%    \end{align*}
%    where $\E{G(y)}$ can be found using \Cref{thm:ev_thm} and \Cref{lem:E_evt_dists}.
%\end{proof}
%%% % % % % % % % % % % % % % % % % % % % % % % % % % % % % % % % % % % % % % %
\begin{proof}[of \Cref{lem:relaunch_vs_not}]
When we keep the original copy, the residual execution time of a straggling task is %satisfies
\begin{align}
\Ykeep &= \min\Set{X_{1:r}, (X | X > T^{(1)})}, \\
\Pr(\Ykeep > x) &= \Pr(X>x)^r \frac{\Pr(X > x + F_X^{-1}(1-p))}{p} \label{eqn:Y_keep_tail}
\end{align}
where $\CondProb{X > x + T^{(1)}}{X > T^{(1)}}$ is the additional time needed for the original copy to finish after forking time $T^{(1)}$. As $n \rightarrow \infty$, $T^{(1)} \rightarrow F_X^{-1}(1-p)$. Thus, the tail distribution of $\Ykeep$ is given by \eqref{eqn:Y_keep_tail}. 

When we kill the original copy, $r+1$ new copies of the straggling task are launched at the forking point. Thus the residual execution time is
\begin{align}
\Ykill &= \min\Set{X_{1:r}, X}, \\
\Pr(\Ykill > x) &= \Pr(X>x)^{r+1}.
\end{align}
Killing the original task is better than keeping it if $\Ykeep$ stochastically dominates $\Ykill$, that is $\Pr(\Ykeep > x) \geq \Pr(\Ykill > x)$ for all $x$. This gives the condition \eqref{eqn:kill_or_keep}. Conversely, keeping the original task is better when the reverse condition holds.
\end{proof}

\begin{proof}[of \Cref{thm:single_fork_sexp}]
\begin{align}
\E{T} &= F_X^{-1}(1-p) + \E{ Y_{pn:pn}} \nonumber ,\\
&= \Delta - \frac{1}{\mu} \ln p + \tilde{a}_{pn} \E{ \GumbelDist} + \tilde{b}_{pn},\\
&= \Delta - \frac{1}{\mu} \ln p + \tilde{a}_{pn} \EMConstantSymbol + \tilde{b}_{pn}.
\end{align}
\begin{align}
\E{\Cost} &= \int_{0}^{1-p} F_X^{-1}(h) dh + p F_X^{-1}(1-p)  + (r+1) p \cdot \E{Y}, \\
&= \int_{0}^{1-p}\left ( \Delta - \frac{1}{\mu} \ln(1-h) \right) dh + p \left(\Delta - \frac{1}{\mu} \ln p \right) \nonumber,\\
& \quad \quad \quad \quad  \quad \quad+  (r+1)p \cdot \E{Y} ,\\
&= \Delta + \frac{1}{\mu} \left( p \ln p + (1-p) \right)  + p \Delta - \frac{p}{\mu} \ln p \nonumber ,\\
&\quad \quad \quad \quad  \quad \quad+ (r+1)p \cdot \E{Y} ,\\
&= \Delta (1 + p) + \frac{1-p}{\mu} + (r+1)p \cdot \E{Y} .
\end{align}
To find $\E{Y}$, $\tilde{a}_{pn}$ and  $\tlb_{pn}$ we consider the cases of relaunching ($l=0$) and no relaunching ($l=1$) separately. 

~\\
\textbf{Case 1: Killing the original task ($\SingleForkKillText$)}
\begin{align}
Y &= \min \Set{ X_1, X_2, \cdots X_{r+1} } \\
&\sim \PSExp{\Delta}{(r+1)\mu} \\
\E{Y} &= \Delta + \frac{1}{(r+1)\mu}
\end{align}
Based on \Cref{thm:domain_of_attractions}, for $ \eta(y) = 1/((r+1)\mu)$ we have
\begin{align}
    \lim_{y \rightarrow \CDUpper{F_Y}} \frac{\fccdf[Y]{y + u \eta(y)}}{\fccdf[Y]{y}}
    &= e^{-u}.
\end{align}
By \Cref{thm:ev_thm} and \Cref{thm:domain_of_attractions}, the maximum of shifted exponential belongs to the Gumbel family with
\begin{align*}
    \tla_{pn} &= \frac{1}{\mu(1+r) },
    \\
    \tlb_{pn} &= \finvccdf[Y]{1/n} = \Delta + \frac{\ln (pn) }{\mu(r+1)}.
\end{align*}

~\\
\textbf{Case 2: Keeping the original task ($\SingleForkKeepText$)}
\label{sec:shifted_exp_no_relaunching}

In the case of no relaunching,
\[Y = \min\Set{\PExp{\mu}, \Delta + \PExp{r\mu}}.\]
Note that the first term does not include $\Delta$ because for large $n$ the original task would have run for at least $\Delta$ seconds. Thus the tail distribution of $Y$ is given by
\begin{align}
    \label{eqn:shifted_exp_Y}
    \fccdf[Y]{y} &= \begin{cases}
        e^{-\mu y} & 0 < y < \Delta, \\
        e^{\mu r \Delta} e^{-\mu(r+1)y} & y \geq \Delta.
    \end{cases}
\end{align}

The expected value $\E{Y}$ is the integration of $\fccdf[Y]{y}$ over its support.
\begin{align*}
\E{Y} &= \int_{0}^{\Delta} e^{-\mu y} dy + \int_{\Delta}^{\infty} e^{\mu r \Delta} e^{-\mu(r+1)y}, \\
&= \frac{1- e^{-\mu \Delta}}{\mu} + \frac{e^{-\mu \Delta}}{ \mu (r+1)} .
\end{align*}

By \Cref{thm:ev_thm} and \Cref{thm:domain_of_attractions} similar to the relaunching case we have
\begin{align*}
    \tla_{pn} &= 1/\left[ \mu(1+r) \right], \\
    \tlb_{pn} &= \finvccdf[Y]{1/n} = \frac{r}{r+1} \Delta + \frac{\ln (pn) }{\mu(r+1)}.
\end{align*}
\end{proof}

%% % % % % % % % % % % % % % % % % % % % % % % % % % % % % % % % % % % % % % %

Before showing the detailed proof of \Cref{thm:single_fork_pareto}], 
we state in \Cref{lem:DA_FY} how the domain of attraction of $F_Y$ relates to that of $F_X$. 

%% % % % % % % % % % % % % % % % % % % % % % % % % % % % % % % % % % % % % % %
\begin{lem}[Domain of attraction for $F_Y$]
    \label{lem:DA_FY}
    Given a single fork policy $\SingleFork{p,r; n}$ with $0 < p < 1$, 
    \begin{enumerate}
        \item if $F_X \in \DAG$, then $F_Y \in \DAG$;
        \item if $F_X \in \DAF$, then $F_Y \in \DAF[(r+1)\xi]$;
        \item if $F_X \in \DAW$, then $F_Y \in \DAW[(r+1)\xi]$ for $\SingleForkKill{p,r}$ and $F_Y \in \DAW[\xi]$ for $\SingleForkKeep{p,r}$.
    \end{enumerate}
\end{lem}
The proof follows directly from \Cref{eqn:Y_defn} and \Cref{thm:domain_of_attractions}, and hence is omitted here.

\begin{proof}[of \Cref{thm:single_fork_pareto}]
From \Cref{thm:single_fork_gen} we have
\begin{align}
\E{T} &= F_X^{-1}(1-p) + \E{ Y_{pn:pn}} \nonumber ,\\
&= x_m p^{-1/\alpha}  + \tilde{a}_{pn} \E{ \Phi_{(r+1)\alpha}} \label{eqn:latency_pareto_1} ,\\
&= x_m p^{-1/\alpha}  + \tilde{a}_{pn}  \Gamma \left( 1- 1\frac{1}{(r+1)\alpha} \right) \label{eqn:latency_pareto_2}. \\
%\end{align}
%\begin{align}
\E{C} &=  \int_{0}^{1-p} F_X^{-1}(h) dh + p F_X^{-1}(1-p)  + (r+1) p \cdot \E{Y} ,\label{eqn:cost_pareto_1} \\
&= x_m \int_{0}^{1-p} (1-h)^{-1/\alpha} dh + p x_m p^{-1/\alpha}  + (r+1) p \cdot \E{Y} ,\nonumber \\
&= x_m \frac{\alpha}{\alpha-1} [1 - p^{1-1/\alpha} ] + x_m p^{1-1/\alpha}  + (r+1) p \cdot \E{Y} ,\nonumber \\
&= x_m \frac{\alpha}{\alpha-1} - x_m \frac{p^{1- 1/\alpha}}{\alpha-1} + (r+1) p \cdot \E{Y} .\label{eqn:cost_pareto_4}
\end{align}
To obtain \eqref{eqn:latency_pareto_1} we first observe that since $F_X$ is Pareto, by
\Cref{thm:domain_of_attractions} it falls into the \Frechet{} domain of attraction, i.e.\ $F_X \in
\DAF[\alpha]$. Then using \Cref{lem:DA_FY} we can show that $F_Y \in \DAF[(r+1)\alpha]$. Subsequently, using
\Cref{thm:ev_thm} and \Cref{lem:E_evt_dists} we get \eqref{eqn:latency_pareto_2}. To derive the expected cost
\eqref{eqn:cost_pareto_4} we substitute $F^{-1}_X(h) = x_m (1-h)^{-1/\alpha}$ in the first and second terms in
\eqref{eqn:cost_pareto_1} and simplify the expression.
To find $\tilde{a}_{pn}$ and $\E{Y}$ in \eqref{eqn:latency_pareto_2} and \eqref{eqn:cost_pareto_4} respectively we
consider the cases of killing the original task ($\SingleForkKillText$) and keeping the original task
($\SingleForkKeepText$) separately. 

~\\
\textbf{Case 1: Killing the original task($\SingleForkKillText$)} \\
For a single-fork policy that kills the original task, the scheduler waits for $(1-p)n$ tasks to finish and then relaunches each of the $pn$ straggler tasks on a new machine. 
\begin{align}
Y &= \min( X_1, X_2, \dots X_{r+1} ) ,\nonumber \\
Y &\sim \PPareto{(r+1)\alpha}{x_m} . \label{eqn:Y_pareto_relaunch}
\end{align}
From \eqref{eqn:a_n_frechet} in \Cref{thm:ev_thm} we can evaluate $\tilde{a}_{pn}$ as follows
\begin{align*}
\tilde{a}_{pn} &= F_Y^{-1}\left(1- \frac{1}{pn} \right) = x_m (pn)^{1/\alpha} .
\end{align*}
And $\E{Y}$ of \eqref{eqn:Y_pareto_relaunch} can be evaluated as
\begin{align}
\E{Y} &= \frac{ (r+1) \alpha }{ (r+1) \alpha - 1} x_m .
\end{align}

~\\
\textbf{Case 2: Keeping the original task ($\SingleForkKeepText$)} \\
For a single-fork policy that keeps the original task, the scheduler keeps the original copy, and adds $r$
additional replicas for each straggling task. Thus the residual execution time can be expressed as
\begin{align}
Y &= \min( \PPareto{\alpha}{F_X^{-1}(1-p)} - F_X^{-1}(1-p),  \PPareto{r\alpha}{x_m} )  \\
\bar{F}_Y(y) &= \frac{1}{p} \left( \frac{x_m}{y} \right)^{\alpha r} \left( \frac{x_m}{y + x_m p^{-1/\alpha}} \right)^{\alpha} .\label{eqn:Y_pareto_no_relaunch}
\end{align}
From \eqref{eqn:a_n_frechet} in \Cref{thm:ev_thm}, $\tilde{a}_{pn} = \bar{F}_Y^{-1}\left(\frac{1}{pn} \right)$. The expected value of $Y$ can be found by numerically integrating $\bar{F}_Y(y)$ in \eqref{eqn:Y_pareto_no_relaunch} over its support. 
%
%which simplifies to
%\begin{align*}
%% (pn)^{1/\alpha} &= \left( 1 + \dfrac{\tla_{pn}}{F_X^{-1}(1-p)} \right)\left( \dfrac{\tla_{pn}}{x_m} \right)^r , \\
% (pn)^{1/\alpha} &= \left( 1 + \dfrac{\tla_{pn}}{x_m p^{-1/\alpha}} \right)\left( \dfrac{\tla_{pn}}{x_m} \right)^r ,
%\end{align*}
%which simplifies to \eqref{eqn:tla_n_for_pareto_no_relaunching}. 
\end{proof}

%% % % % % % % % % % % % % % % % % % % % % % % % % % % % % % % % % % % % % % %
\begin{proof}[of \Cref{coro:scaling}]
For the case of killing the original task, it follows directly from \eqref{eqn:pareto_ET} and
\eqref{eqn:pareto_relaunching_a}. For the case of keeping the original task, note that $\tla_{pn}$ grows with $n$. When $n$ is large enough, from\eqref{eqn:Y_pareto_no_relaunch} and the fact that $\tilde{a}_{pn} = \bar{F}_Y^{-1}\left(\frac{1}{pn} \right)$, we have
\begin{align}
\tla_{pn} ^{r+1}
\leq
n^{1/\alpha} x_m^{r+1} 
&\leq 2 \tla_{pn} ^{r+1},
\end{align}
and then the result holds again following \eqref{eqn:pareto_ET}.
\end{proof}

\bibliographystyle{plain}
\bibliography{bibtex/computing,bibtex/storage}

\end{document}